\documentclass{aa}  

\usepackage{graphicx}
\usepackage{subcaption}
\usepackage{txfonts}
\usepackage{hyperref}

\usepackage[normalem]{ulem}
\usepackage[x11names]{xcolor}
\usepackage{academicons}
\definecolor{orcidlogocol}{HTML}{A6CE39}

\newcommand{\orcid}[1]{\href{https://orcid.org/#1}{\textcolor[HTML]{A6CE39}{\aiOrcid}}}
\begin{document}

   \title{Thermally enhanced tearing in solar current sheets: explosive reconnection with plasmoid-trapped condensations}


   \author{Samrat Sen \href{https://orcid.org/0000-0003-1546-381X}{\includegraphics[scale=0.05]{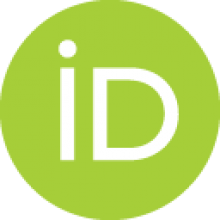}}
          \inst{1}$^*$
          \and
          Rony Keppens\href{https://orcid.org/ 0000-0003-3544-2733 }{\includegraphics[scale=0.05]{orcid-ID.png}} \inst{1}
          }

   \institute{$^1$Centre for mathematical Plasma-Astrophysics, Celestijnenlaan 200B, 3001 Leuven, KU Leuven, Belgium\\
              $^*$\email{samrat.sen@kuleuven.be; samratseniitmadras@gmail.com}
                      }

   \date{Received: XXXX; accepted: XXXX}
   \date{}
 
  \abstract
   {Thermal instability plays a major role in condensation phenomena in the solar corona, e.g. for coronal rain and prominence formation. In flare-relevant current sheets, tearing instability may trigger explosive reconnection and plasmoid formation. However, how both instabilities influence the disruption of current concentrations in the solar corona has received less attention to date.}
   {We explore how the thermal and tearing modes reinforce each other in the fragmentation of a current sheet in the solar corona through an explosive reconnection process, characterized by the formation of plasmoids which interact and trap condensing plasma.}
   {We use a resistive magnetohydrodynamic (MHD) simulation of a 2D current layer, incorporating the non-adiabatic effects of optically thin radiative energy loss and background heating using the open-source code \texttt{MPI-AMRVAC}. Multiple levels of adaptive mesh refined grids are used for achieving a high resolution to resolve the fine structures during the evolution of the system.}
   { 
   Our parametric survey explores different resistivities and plasma-$\beta$ to quantify the instability growth rate in the linear and nonlinear regimes. We notice that for dimensionless resistivity values within $10^{-4} - 5 \times 10^{-3}$, we get explosive behavior where thermal instability and tearing behavior reinforce each other. This is clearly below the usual critical Lundquist number range of pure resistive explosive plasmoid formation. We calculate the mean growth rate for the linear phase and different non-linear phases of the evolution. The non-linear growth rates follow weak power-law dependency with resistivity. The fragmentation of the current sheet and the formation of the plasmoids in the nonlinear phase of the evolution due to the thermal and tearing instabilities are obtained. The formation of plasmoids is noticed for the Lundquist number ($S_L$) range $4.6 \times 10^3 - 2.34 \times 10^5$. We quantify the temporal variation of the plasmoid numbers and the density filling factor of the plasmoids for different physical conditions. We also find that the maximum plasmoid numbers scale as $S_L^{0.223}$.Within the nonlinearly coalescing plasmoid chains, localized cool condensations gather, realizing density and temperature contrasts similar to coronal rain or prominences.}
   {}


\titlerunning{Thermal and tearing modes in a current sheet}
\authorrunning{S.Sen \& R.Keppens}

\maketitle


\section{Introduction} 

Magnetic reconnection is ubiquitous in both laboratory and astrophysical plasmas, where the change of magnetic field topology leads to conversion of the magnetic energy into thermal and kinetic energies \citep{2000mrp..book.....B}. Magnetic reconnection plays a key role in the fast energy release in solar flares \citep{1939ApJ....89..555G, 1947MNRAS.107..338G, 1948MNRAS.108..163G, 2000mare.book.....P, 2020JGRA..12525935H}, in coronal mass ejections into the ambient solar wind medium \citep{1995GeoRL..22.1753G, 2003JGRA..108.1023S, 2012ApJ...760...81K}, and mediates the loss of plasma confinement in tokamak experiments \citep{2015PPCF...57a4017G}. The rearrangement of the magnetic field topology occurs in a localized plasma region where non-ideal magnetohydrodynamics (MHD) effects dominate, breaking the frozen-in condition. The Sweet-Parker model \citep{1957JGR....62..509P, 1958IAUS....6..123S} predicts the reconnection rate to scale with the Lundquist number ($S_L= L v_A/\eta$) as $S_L^{-1/2}$, where $L$ is the characteristic length, $v_A$ is the Alfv{\'e}n velocity, and $\eta$ is the resistivity of the medium. However, this prediction is too slow to agree with reconnection observations for the solar atmosphere. On the other hand, the Petschek model \citep{1964NASSP..50..425P} estimates the reconnection rate to scale as $(\mathrm{log}\ S_L)^{-1}$. \cite{2000mrp..book.....B} and \cite{2010PhPl...17f2104H} state that the Petschek reconnection is achievable only if the local resistivity of the current sheet is enhanced, while \cite{2009PhPl...16a2102B} reports the occurrence of Petscheck like reconnection for a low uniform resistivity ($10^{-3}$). 

The simplest configuration susceptible to magnetic reconnection is a single current layer model formed by a polarity reversal of the magnetic field. This reconnection may be triggered due to the growth of a classical linear resistive instability, known as tearing mode instability \citep{1963PhFl....6..459F}. A current sheet of aspect ratio $L/\delta \gtrsim 2\pi$ (where $L$ and $\delta$ are the characteristic length and thickness of the current layer respectively) can develop magnetic islands due to the growth of linearly unstable perturbations. \cite{2014ApJ...780L..19P, 2015ApJ...806..131L, 2016JPlPh..82e5301T} have reported the development of ideal tearing modes in current sheets for a large aspect ratio of $\sim S_L^{1/3}$. On the other hand, double current layer models are also seen to give rise to resistive instabilities known as double tearing modes (DTMs). Double current layer models with two widely seperated current layers can develop a single standard tearing mode on each layer, to influence each other later in the nonlinear evolution stage \citep{2013PhPl...20i2109K, 2021PhPl...28h2903P}. In contrast, DTMs are tearing modes that are intrinsically coupled and that co-develop on nearby resonant surfaces. The evolution of DTMs in the nonlinear regime has been seen to lead to an explosive reconnection and a weak dependence on the resistivity \citep{2011PhPl...18e2303Z, 2017PhPl...24h2116A} (and references therein). DTMs have been studied under various important conditions like external shear flows \citep{1992PhFlB...4.2751O, 2007PhPl...14a0704B, 2008PhPl...15h2109W}, bootstrap current \citep{1997PhPl....4.1047Y}, anomalous electron viscosity \citep{2003PhPl...10.3151D},
collisionless plasma \citep{2007PPCF...49..675B}, and Hall effects \citep{2008PlST...10..407Z, 2009PhPl...16l2113Z}. \cite{2018PhPl...25c2113P} estimated scaling relations between the maximum growth rate, Lundquist number and the aspect ratio, for linear tearing modes in a double current sheet set up.   

Thermal instability is an essential mechanism to form condensations in the solar atmosphere. The theory is laid out in classical treatments by \cite{1953ApJ...117..431P} and \cite{1965ApJ...142..531F}. These works explain how a runaway process of radiative cooling leads to thermal instability (TI) in plasma. The solar corona may be considered to be in a delicate thermal equilibrium balancing the optically thin radiative loss and background heating in combination with thermal conduction. If this balance is perturbed, and the increment of radiative loss cools down the plasma, isobaric, isentropic or isochoric evolutions may self-amplify the radiative losses. This drives the enhancement of the local plasma density, which further increases the energy loss by radiation (because radiative energy loss in optically thin medium varies with the density squared), which in turn drops the temperature even more. Hence, a catastrophic runaway process results in a rapid drop in temperature and an increase in plasma density. \cite{1965ApJ...142..531F} reported a detailed analysis of the thermal instability in an infinite homogeneous medium which triggers the catastrophic radiative cooling. Later, the analysis was extended to non-uniform slab geometry \citep{1991SoPh..131...79V, 1992SoPh..140..317V}, and cylindrical flux tubes with solar coronal conditions \citep{1991SoPh..134..247V, 1995SoPh..160..303I, 2011ApJ...731...39S}. The thermal instability theory can be invoked to explain various fascinating features of the solar atmosphere. E.g., \cite{1977SoPh...53...25S} discussed the formation of a solar prominence in a current sheet, and \cite{1979SoPh...64..267P} extended the study for prominence formation to solar coronal arcades. \cite{1991SoPh..135..361F} used 2D MHD simulation in a line-tied current sheet with the effect of radiative energy losses to explain the formation of post-flare loops. The ab-initio formation of a solar prominence due to chromospheric evaporation and thermal instability is shown by \cite{2012ApJ...748L..26X} in a 2.5D simulation, and the dynamical evolution of similar prominence setups is further explored in \cite{2014ApJ...789...22K}. Later, \cite{2016ApJ...823...22X} developed a 3D model of prominence formation due to the plasma cycle between corona and chromosphere, while the complex 3D dynamics in a twin-layer prominence is reported by \cite{2016ApJ...825L..29X}. More recently, linear and non-linear stability analysis of thermal instability due to the interaction of the entropy and slow MHD wave modes is discussed in \cite{2019A&A...624A..96C, 2020A&A...636A.112C}, and the effect on the thermal instabilities arising from different radiative cooling models is reported by \cite{2021A&A...655A..36H}. The formation of fine structures in the prominence may well relate to the linear magnetothermal modes affected by anisotropic thermal conduction with finite (albeit small) conduction across the magnetic field lines \citep{1991SoPh..134..247V}. Recent solar applications include the works by \cite{2022ApJ...926..216L}, who studied the formation of coronal rain due to the thermal instability of randomly heated arcades, or the formation of prominences due to levitation-condensation by \cite{2021A&A...646A.134J}, as well as a novel plasmoid-fed prominence formation scenario during flux rope eruption by \cite{2022ApJ...928...45Z}. In the latter work, chromospheric plasma collects into a current sheet, which ultimately shows chaotic plasmoid formation, where the cool chromospheric plasma gets trapped and lifted into an erupting prominence structure. This study motivates our current work, where we will investigate more rigorously how thermal and tearing effects can reinforce eachother.

In this work, we study the tearing and thermal instabilities of a single current sheet model with the non-adiabatic effects of radiative energy loss and background heating in a resistive 2D MHD simulation. Recent theoretical studies by \cite{2021SoPh..296...74L, 2021SoPh..296...93L, 2021SoPh..296..117L} show that the instability growth rate in the linear regime of a tearing mode is modified when the non-adiabatic effects, radiative energy loss,  electrical and thermal conductivities are incorporated. This motivates us to explore the growth rate of a thermally influenced tearing mode in the linear and non-linear domains by means of MHD simulation. The focus of our idealized study is mainly devoted to estimating the characteristic growth rate time scales in the linear and non-linear regimes and finding scaling relations with the resistivity in the different phases of the non-linear evolution. We also quantify the number of  generated plasmoids as influenced by different physical parameters, and its scaling relation with Lundquist number. In various earlier works, \cite{2011PhPl...18e2303Z, 2013PhPl...20i2109K,2017PhPl...24h2116A, 2021PhPl...28h2903P} have studied the tearing mode effect in explosive reconnection events for adiabatic conditions. By including the radiative energy loss in an optically thin medium and background heating to investigate the effect of thermal instability in a current sheet model, we determine whether explosive reconnection may be triggered in different Lundquist regimes than in pure resistive MHD alone.

The rest of the paper is organized as follows. In section \ref{setup}, we describe the model setup and the numerical framework along with the initial and boundary conditions. In section \ref{results}, the main results of the study and its analysis are reported. Section \ref{discussion} discusses the significance of the work for a typical coronal medium, summarizes the key findings, and finally concludes how our ﬁndings may be useful for future studies.  



\begin{figure}[hbt!]
    \centering
    \includegraphics[width=0.4\textwidth]{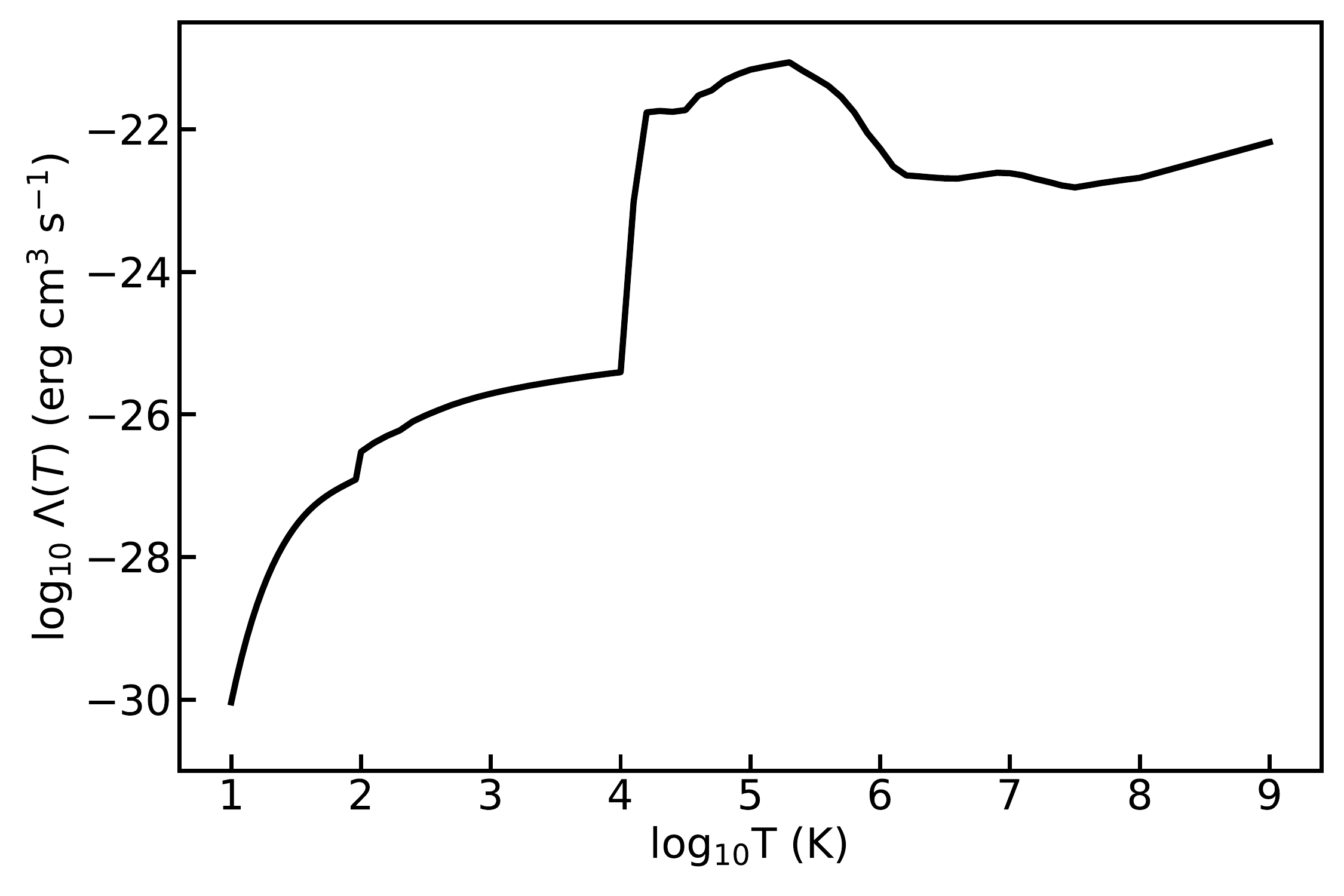}
    \caption{Radiative cooling curve due to `Colgan\_DM' model.}
    \label{fig:clogan_dm}
\end{figure}

\newpage

\begin{figure*}[hbt!]
\centering
\begin{subfigure}{0.42\textwidth}
    \includegraphics[width=1.1\textwidth]{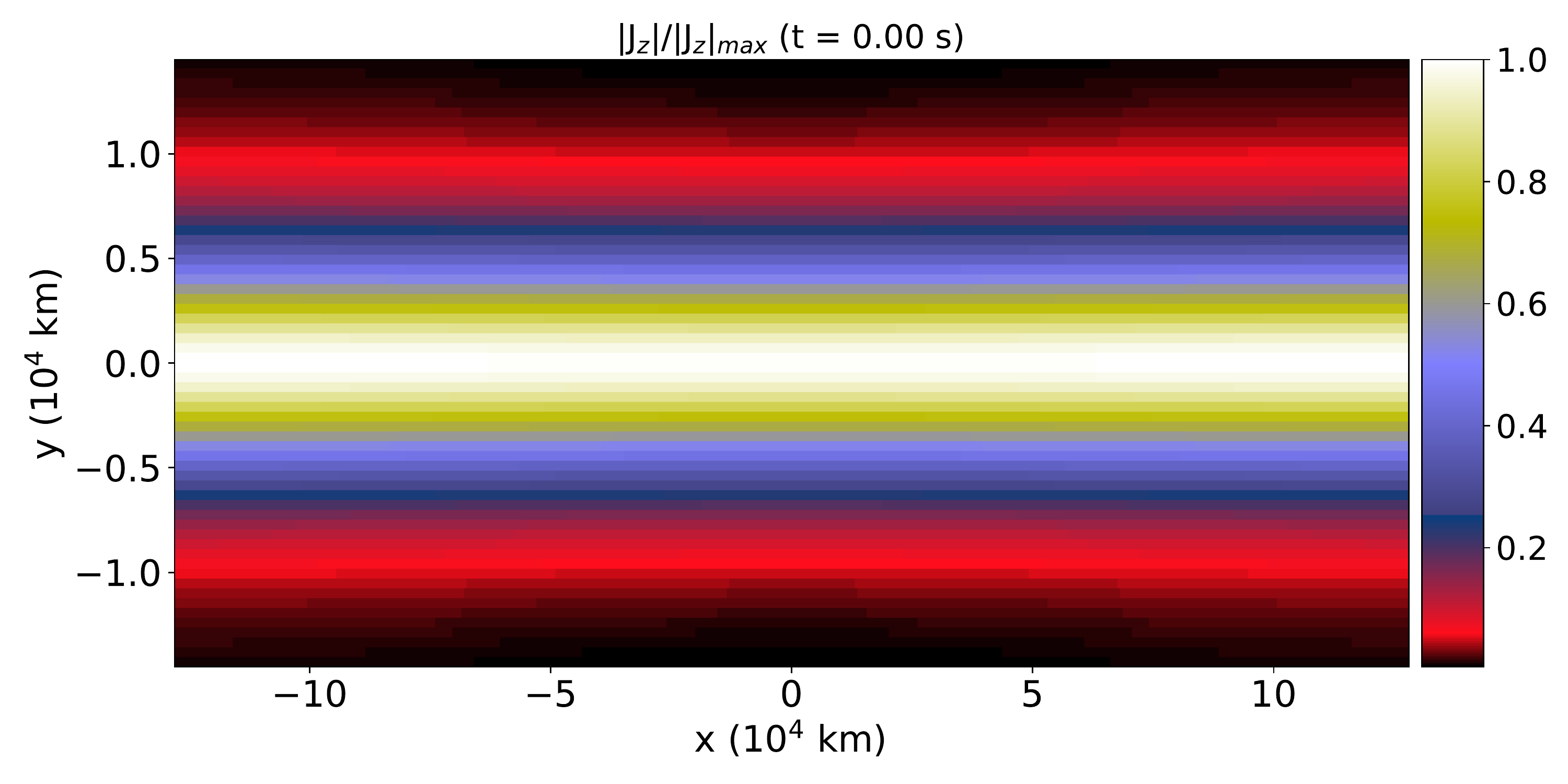}
    \caption{}
    \label{fig:jz_t00}
\end{subfigure}
\hspace{1.5 cm}
\begin{subfigure}{0.42\textwidth}
    \includegraphics[width=1.1\textwidth]{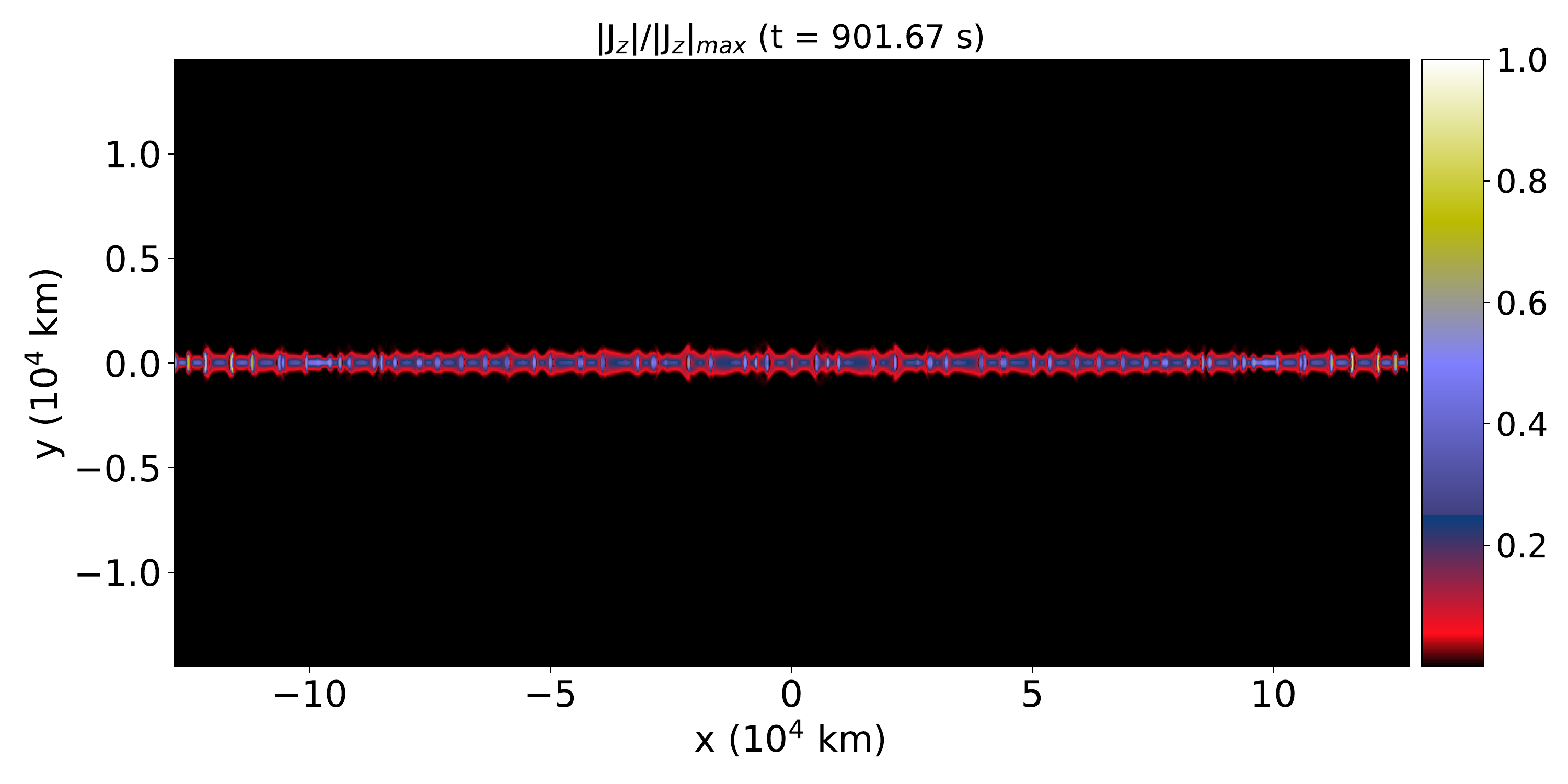}
    \caption{}
    \label{fig:jz_t21}
\end{subfigure}
\newline
\begin{subfigure}{0.42\textwidth}
    \includegraphics[width=1.1\textwidth]{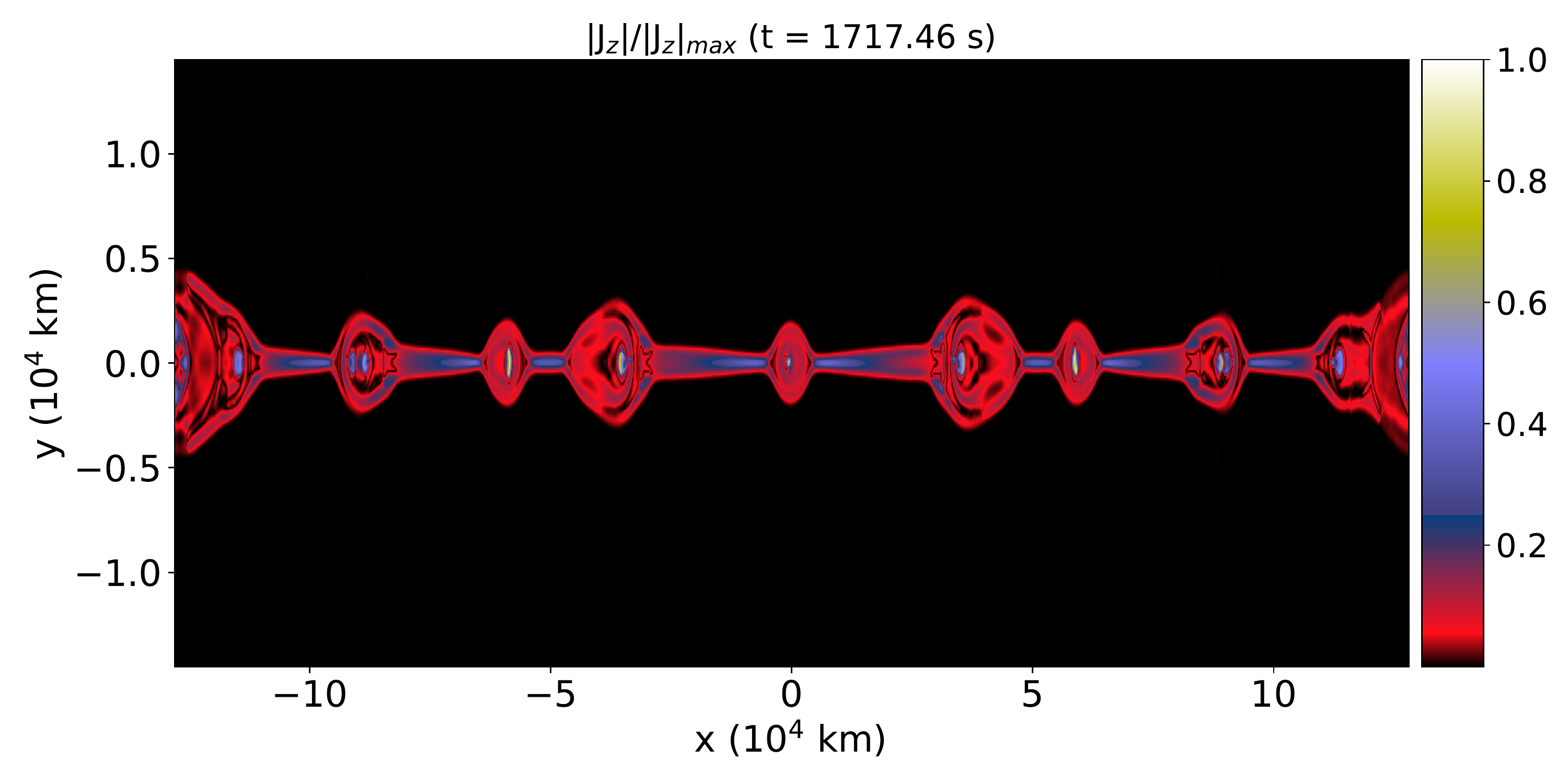}
    \caption{}
    \label{fig:jz_t40}
\end{subfigure}
\hspace{1.5 cm}
\begin{subfigure}{0.42\textwidth}
    \includegraphics[width=1.1\textwidth]{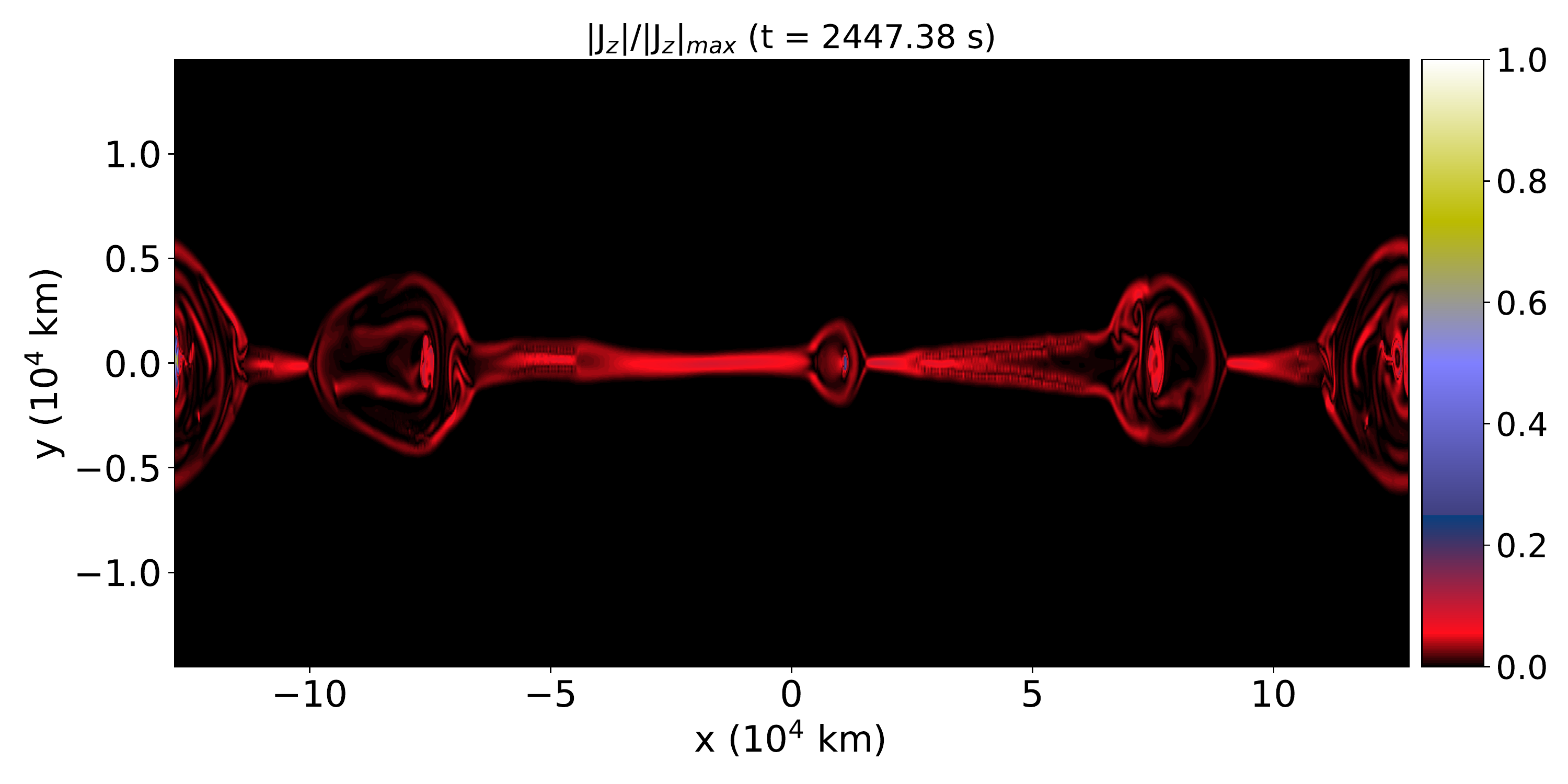}
    \caption{}
    \label{fig:jz_t57}
\end{subfigure}
\caption{Distribution of the absolute current density, $|J_z|$ normalized with the instantaneous absolute peak current density, $|J_{max}|$ for different evolution stages. The $y$-domain is only shown between $-1.45 \times 10^4\ \mathrm{km} \leq y \leq +1.45 \times 10^4\ \mathrm{km}$, which contains the region of the current sheet. (An animation of the figures is available online).}
\label{fig:jzmap}
\end{figure*}

\begin{figure}[hbt!]
\centering
    \includegraphics[width=0.45\textwidth]{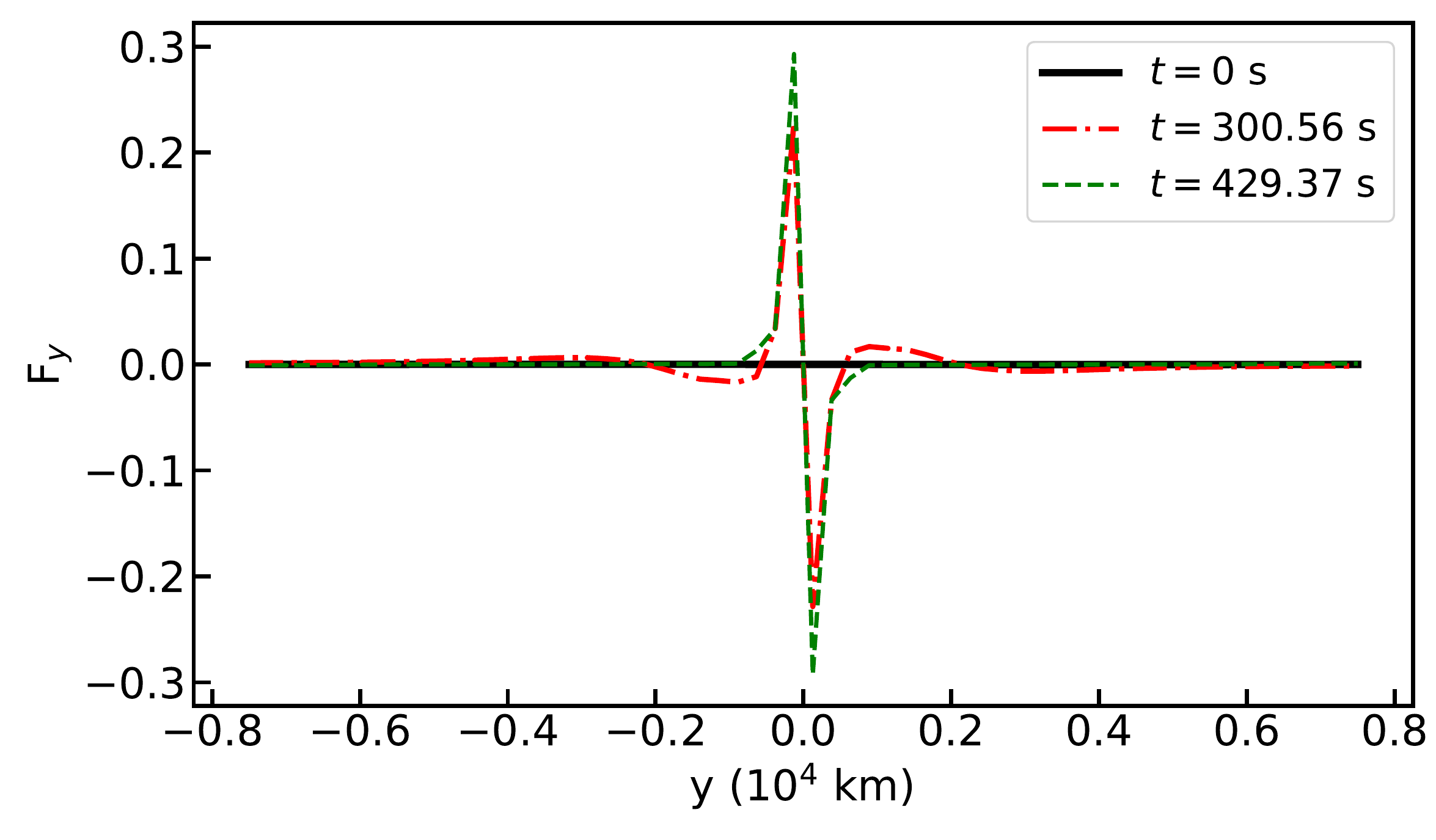}
    \caption{Variation of $\displaystyle{F_y = -\frac{\partial p }{\partial y}+B_x \bigg(\frac{\partial B_y}{\partial x}-\frac{\partial B_x}{\partial y}\bigg)}$ along the $y$-direction between $y=\pm 0.75 \times 10^4$ km for three different times ($t=0$, 300 and 429 s) before the fragmentation stage of the current sheet.}
    \label{fig:Fy}
\end{figure}

\begin{figure*}[hbt!]
\centering
\begin{subfigure}{0.42\textwidth}
    \includegraphics[width=1.1\textwidth]{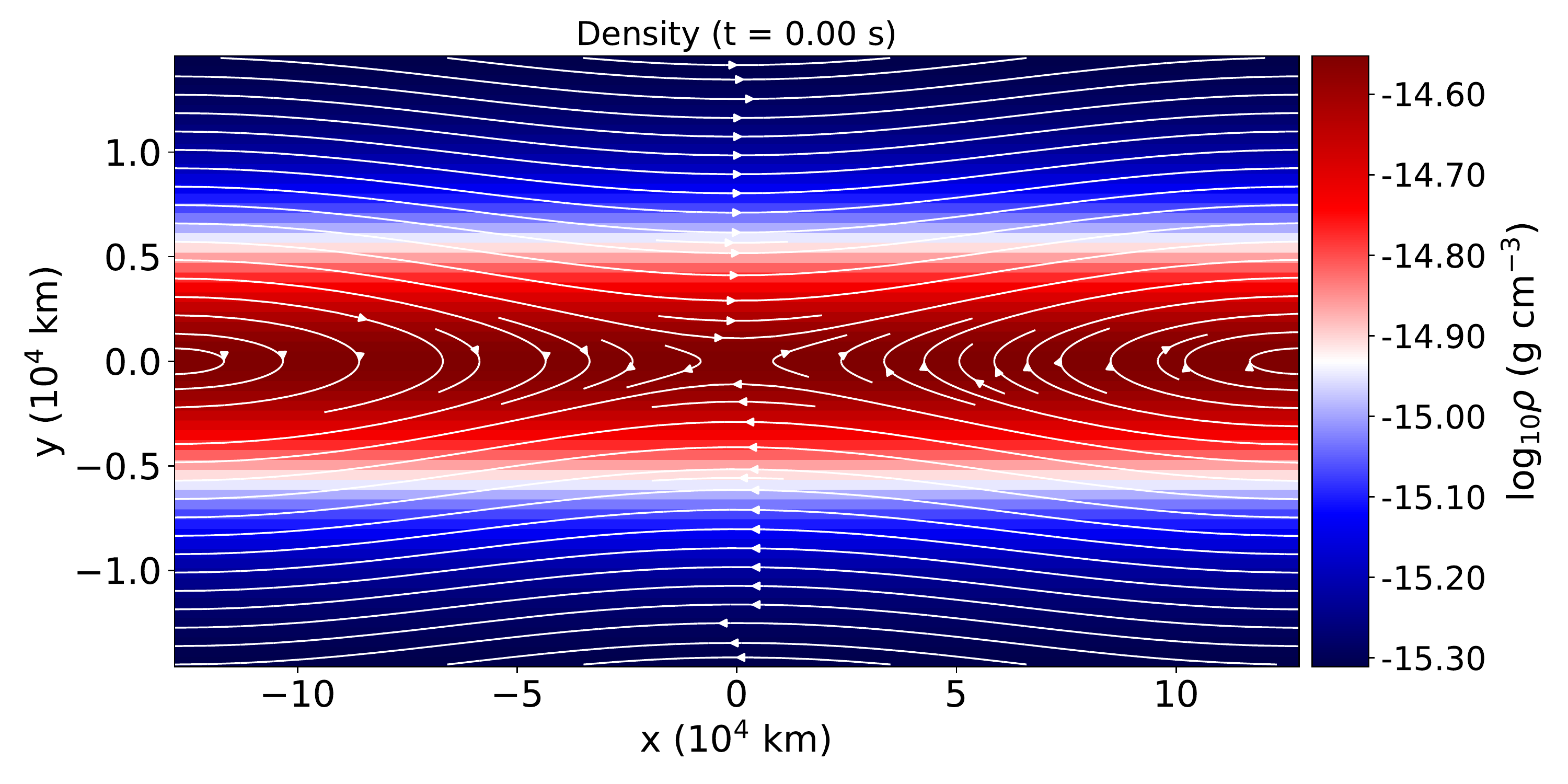}
    \caption{}
    \label{fig:rhomap_t00}
\end{subfigure}
\hspace{1.5 cm}
\begin{subfigure}{0.42\textwidth}
    \includegraphics[width=1.1\textwidth]{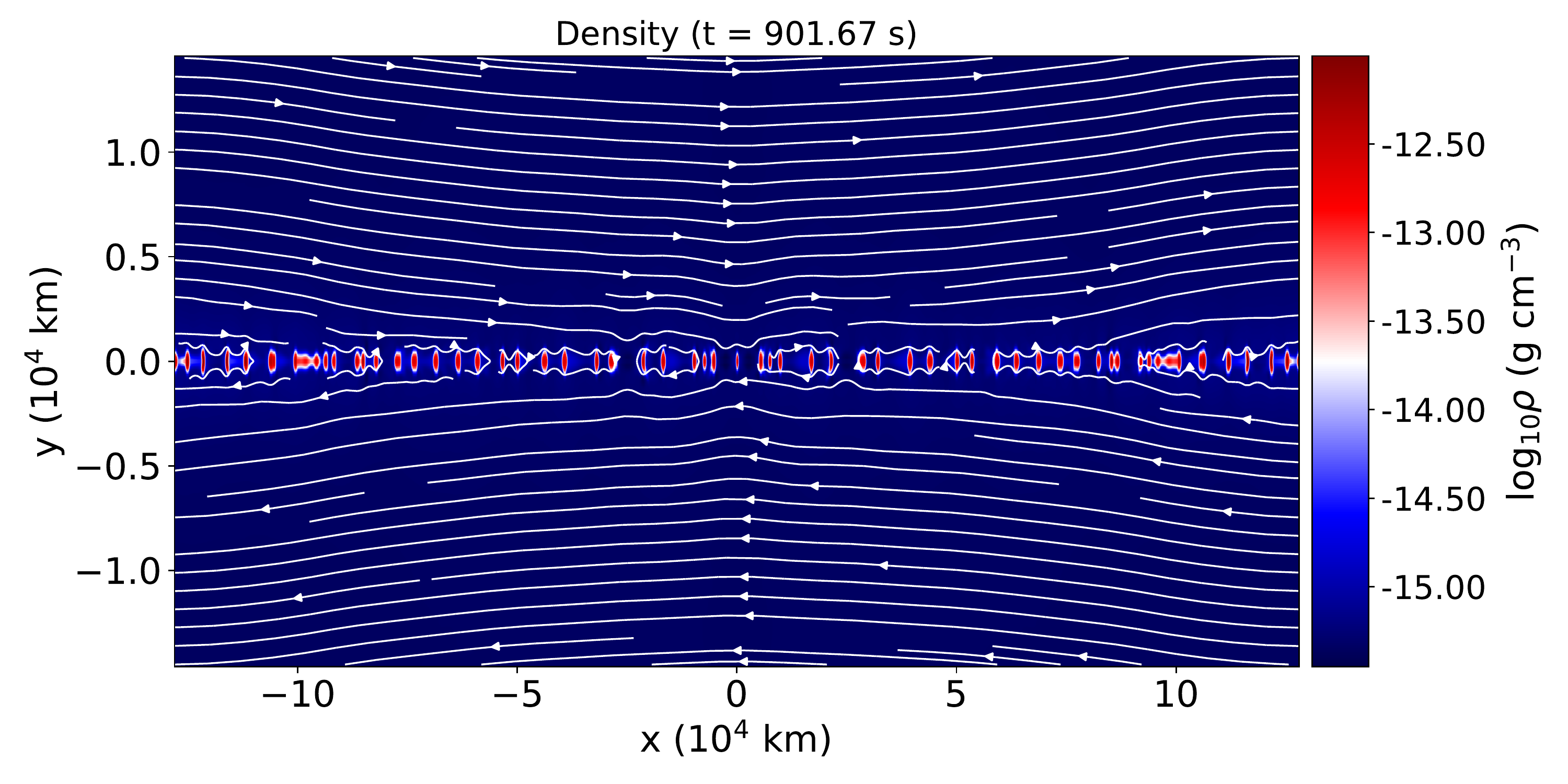}
    \caption{}
    \label{fig:rhomap_t21}
\end{subfigure}
\newline
\begin{subfigure}{0.42\textwidth}
    \includegraphics[width=1.1\textwidth]{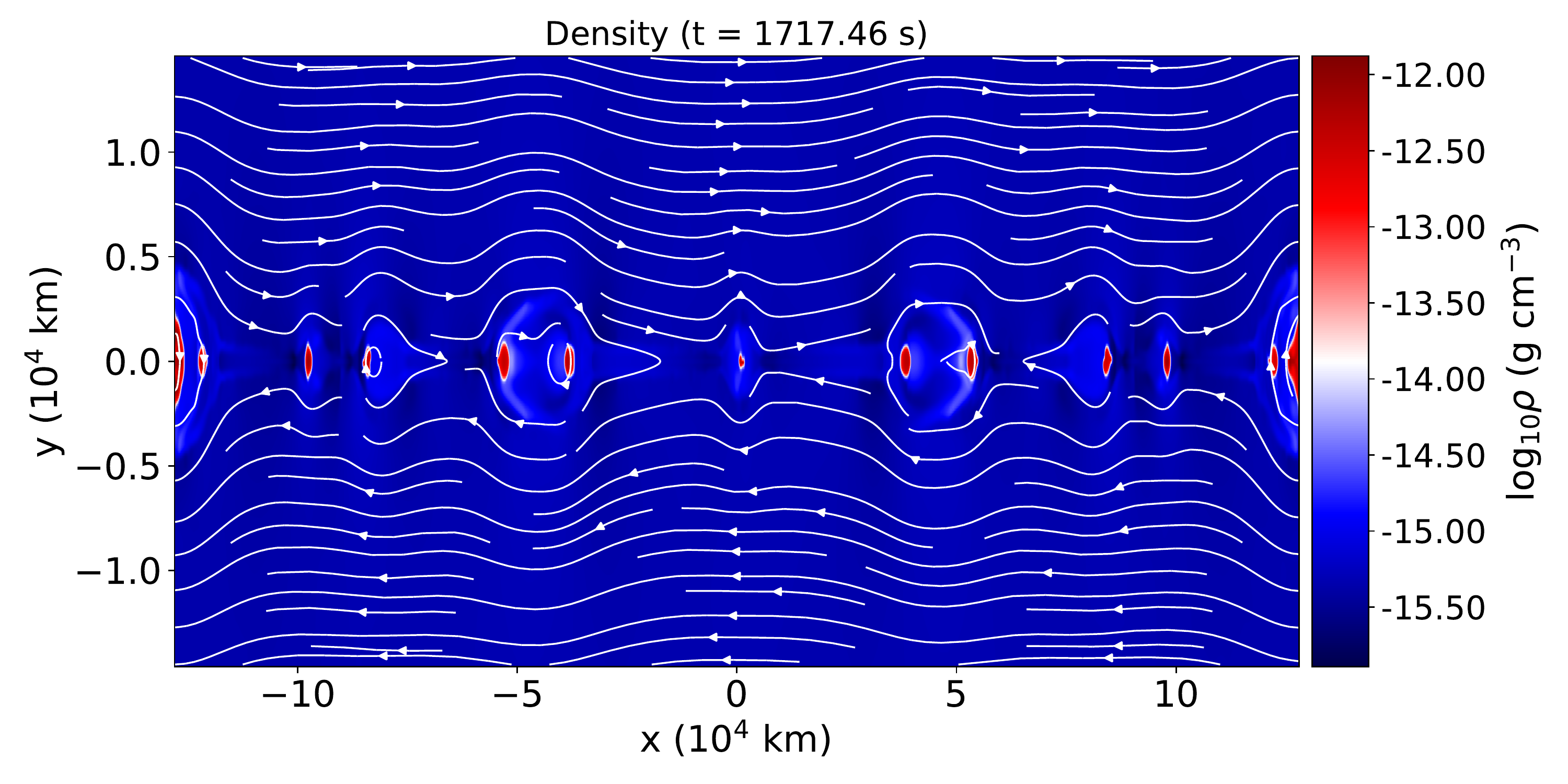}
    \caption{}
    \label{fig:rhomap_t40}
\end{subfigure}
\hspace{1.5 cm}
\begin{subfigure}{0.42\textwidth}
    \includegraphics[width=1.1\textwidth]{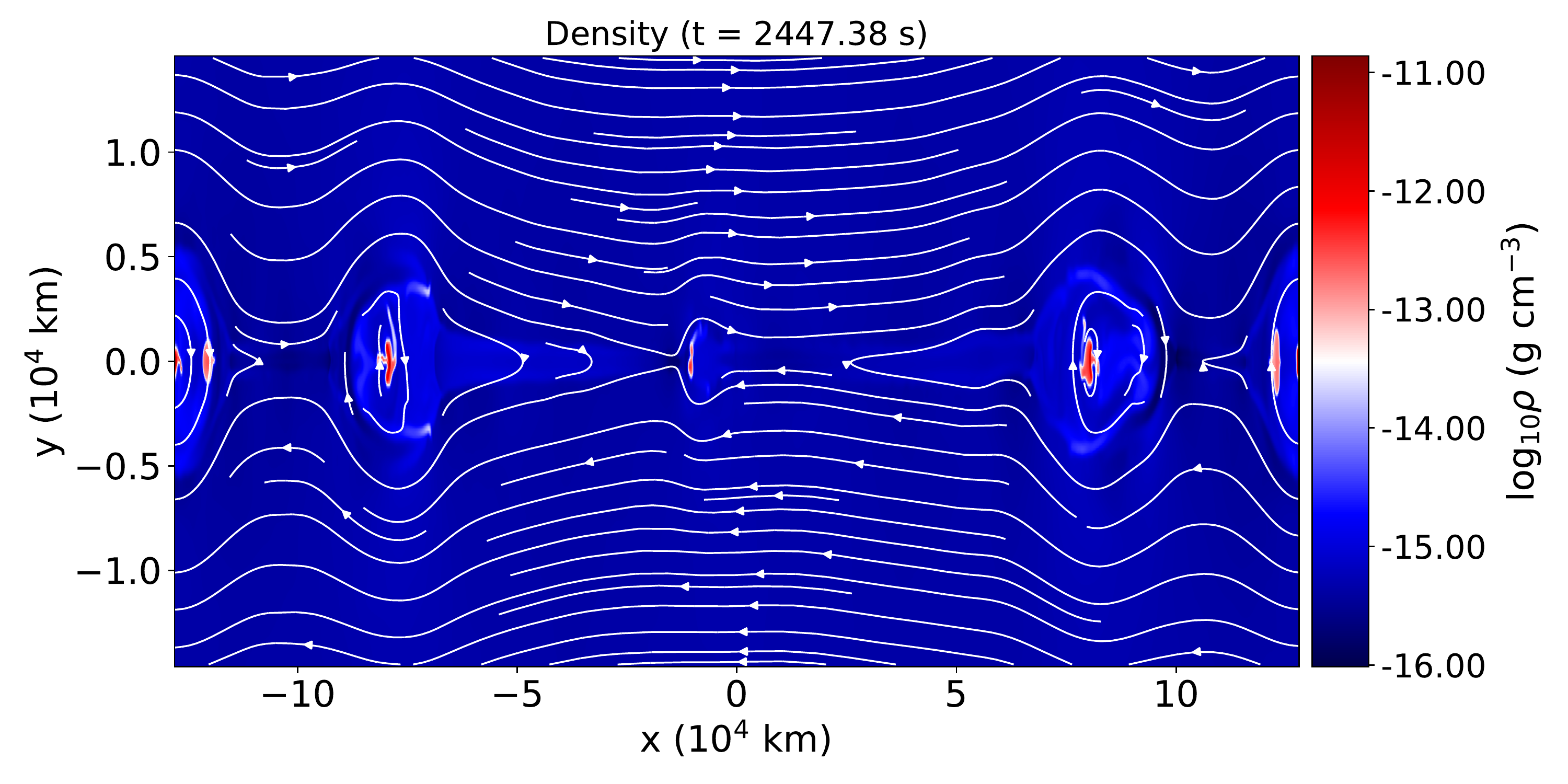}
    \caption{}
    \label{fig:rhomap_t57}
\end{subfigure}
\caption{Same as Fig. \ref{fig:jzmap} for plasma density, $\rho$ for different evolution stages. The over-plotted white lines represent the magnetic field lines. (An animation of the figures is available online).}
\label{fig:rhomap_wfl}
\end{figure*}

\begin{figure*}[hbt!]
\centering
\begin{subfigure}{0.4\textwidth}
    \includegraphics[width=1.1\textwidth]{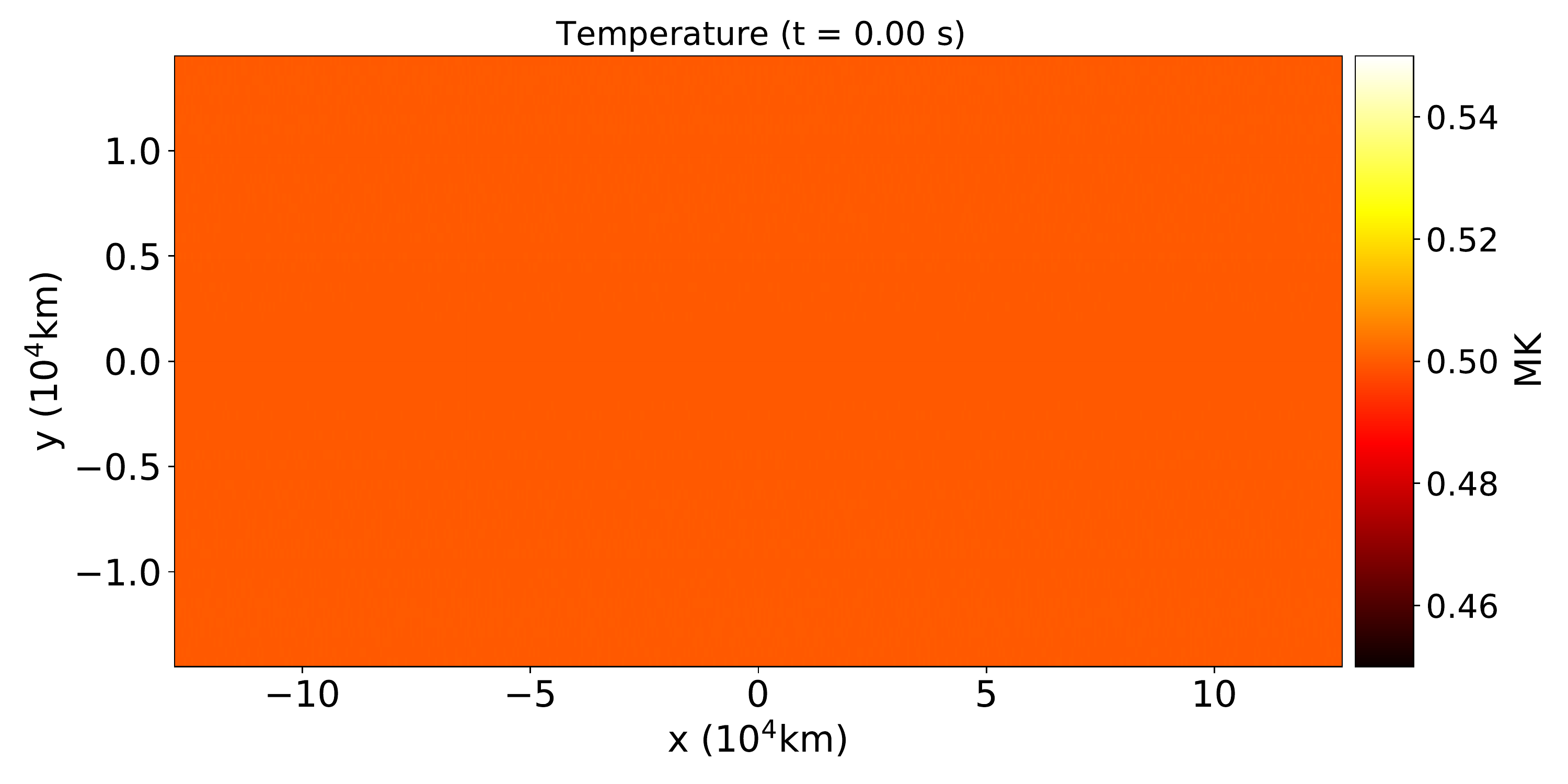}
    \caption{}
    \label{fig:Tmap_t00}
\end{subfigure}
\hspace{1.5 cm}
\begin{subfigure}{0.4\textwidth}
    \includegraphics[width=1.1\textwidth]{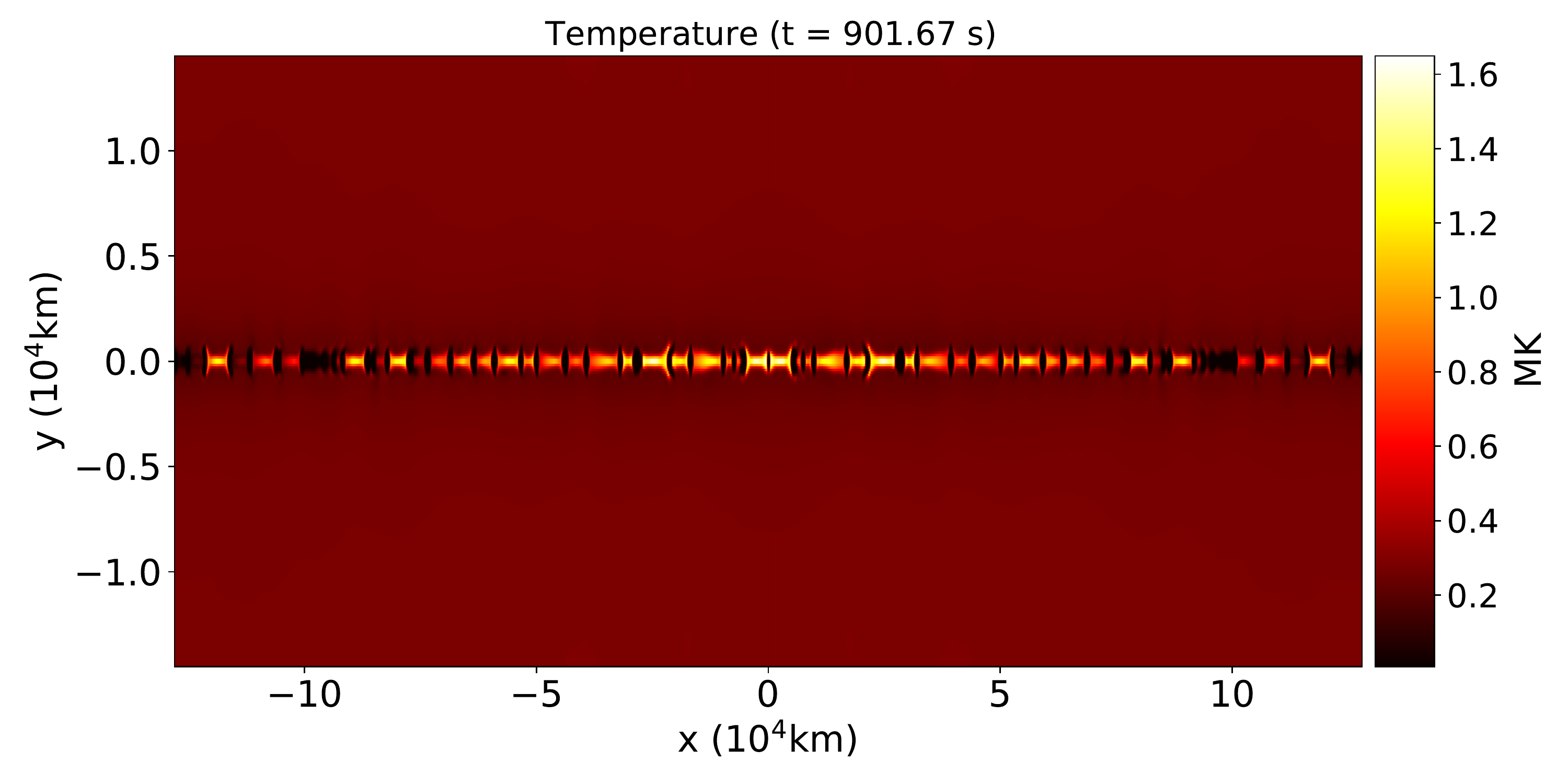}
    \caption{}
    \label{fig:Tmap_t21}
\end{subfigure}
\hspace{1.5 cm}
\begin{subfigure}{0.4\textwidth}
    \includegraphics[width=1.1\textwidth]{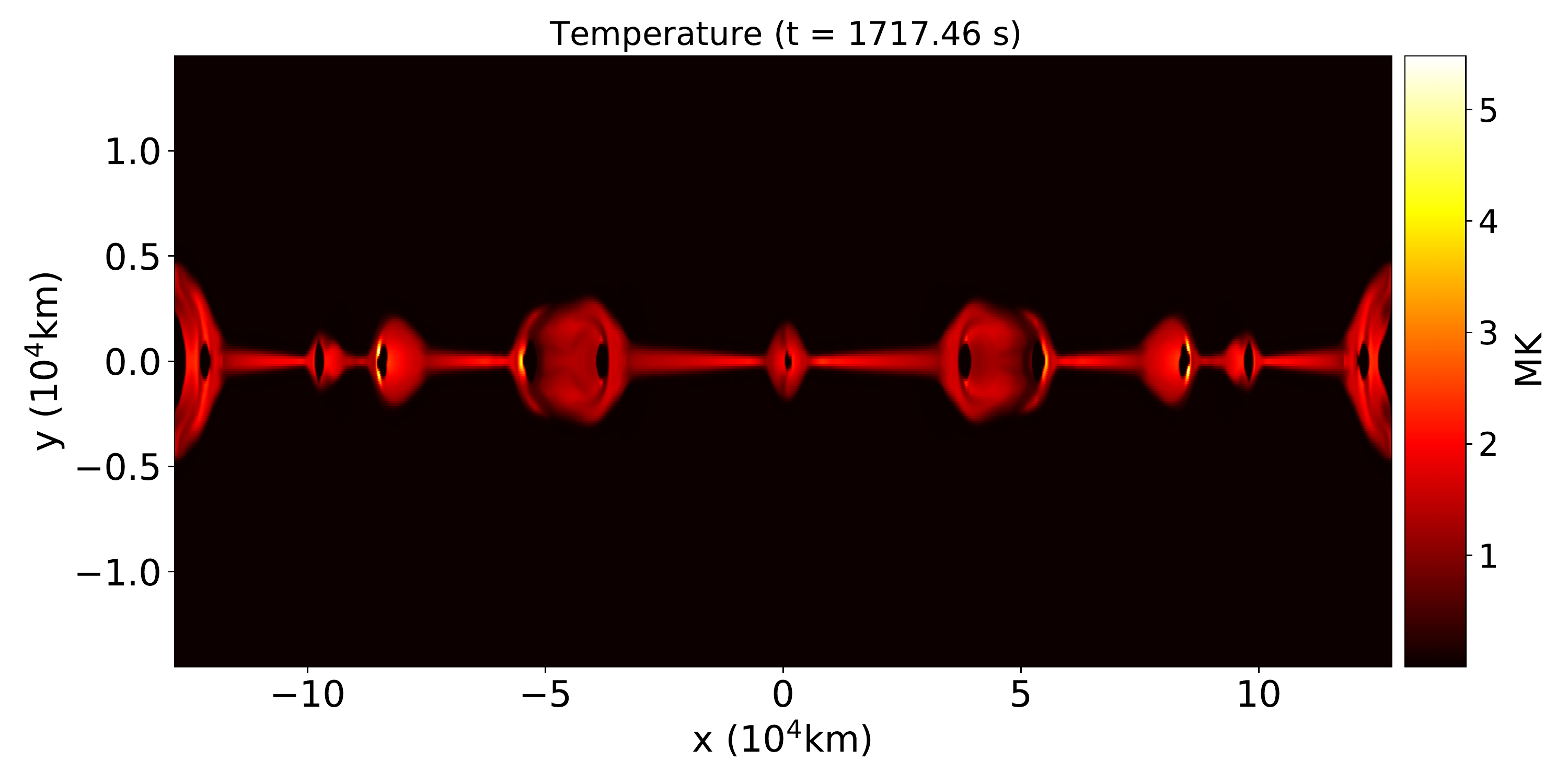}
    \caption{}
    \label{fig:Tmap_t40}
\end{subfigure}
\hspace{1.5 cm}
\begin{subfigure}{0.4\textwidth}
    \includegraphics[width=1.1\textwidth]{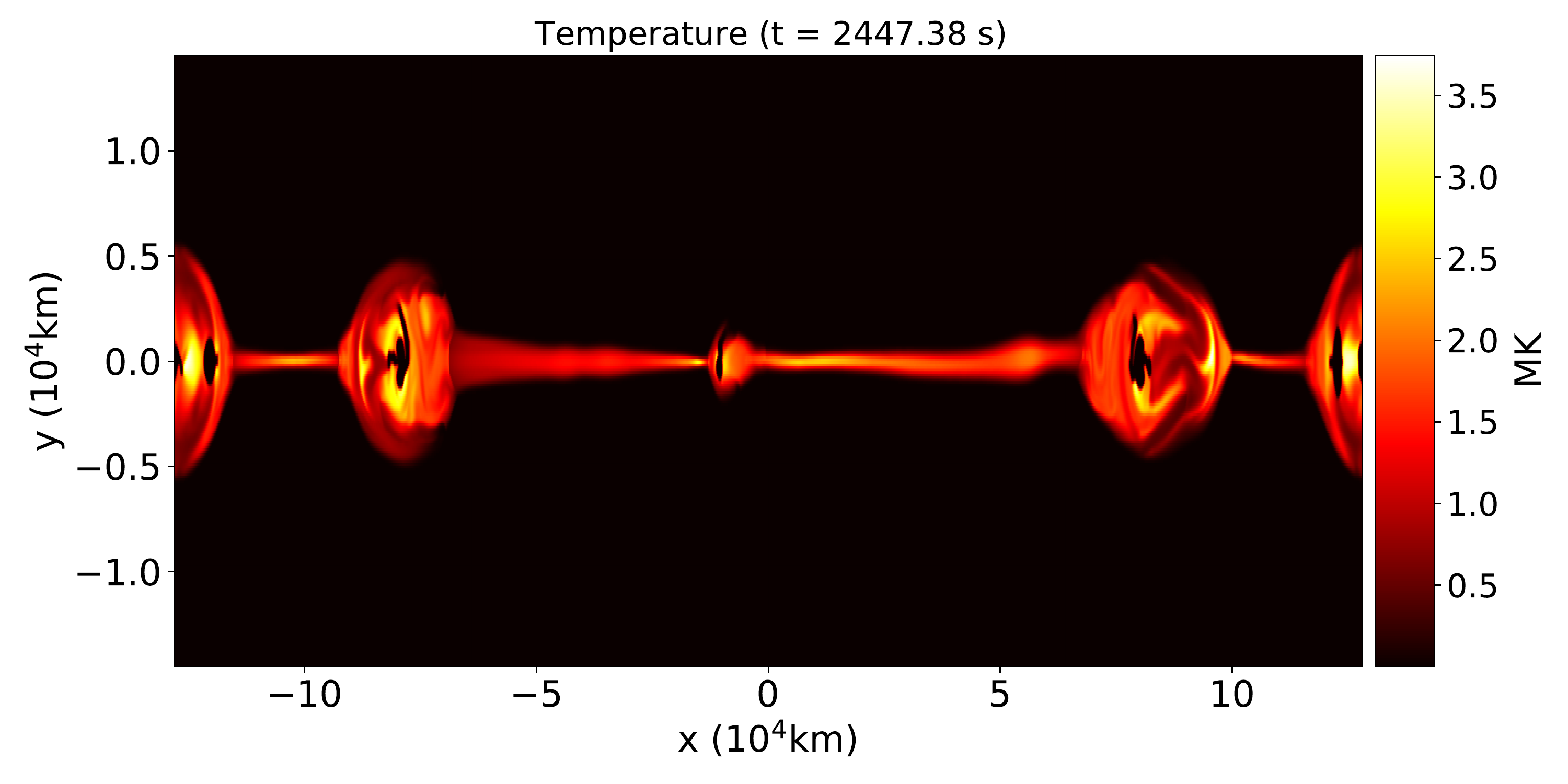}
    \caption{}
    \label{fig:Tmap_t57}
\end{subfigure}
\caption{Distribution of the temperature, $T$ within the same domain as Fig. \ref{fig:jzmap}, for different evolution stages. (An animation of the figures is available online).}
\label{fig:Tmap}
\end{figure*}

\begin{figure*}[hbt!]
\centering
\begin{subfigure}{0.4\textwidth}
    \includegraphics[width=1.1\textwidth]{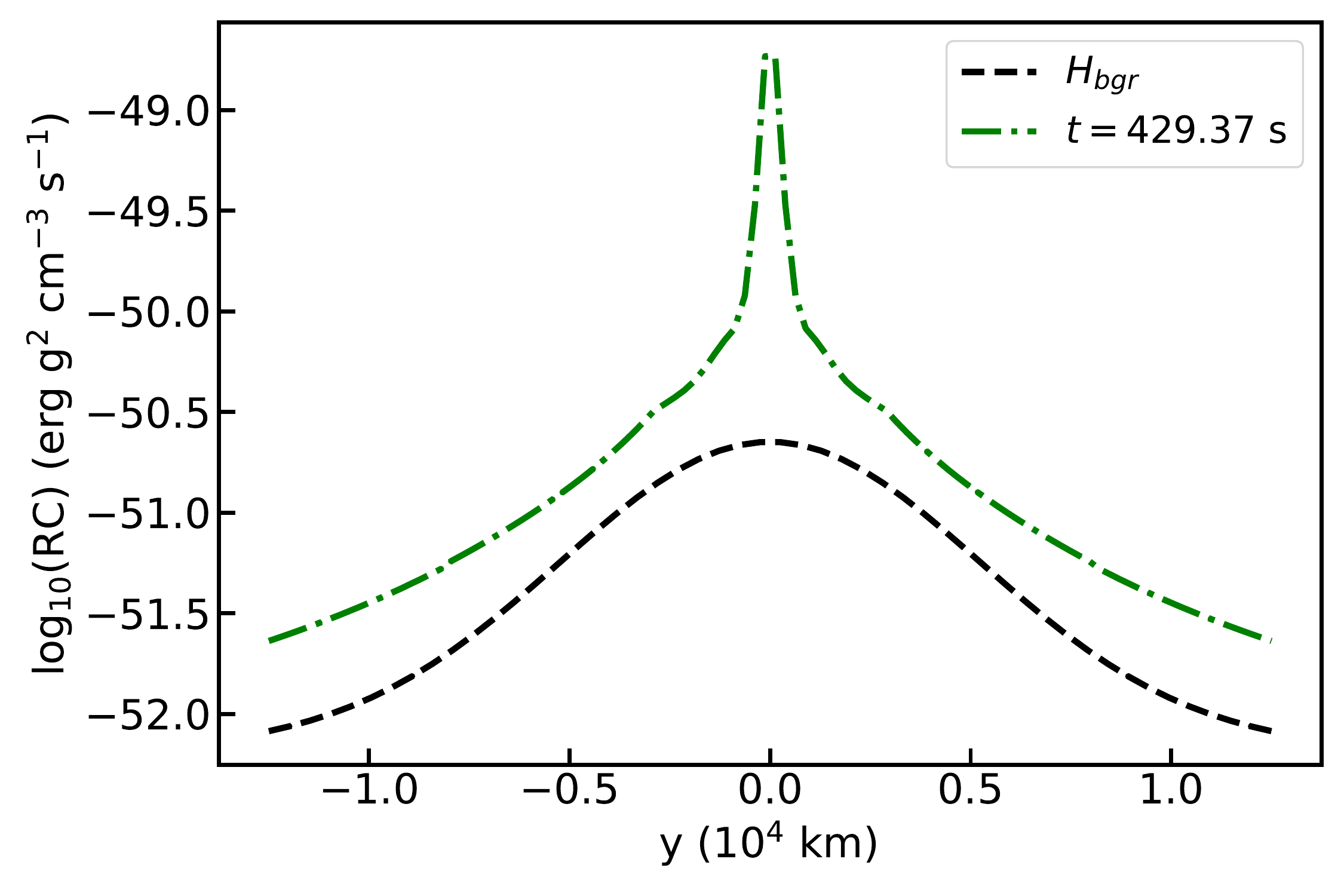}
    \caption{}
    \label{fig:rc_ycut}
\end{subfigure}
 \hspace{1 cm}
\begin{subfigure}{0.43\textwidth}
    \includegraphics[width=1.1\textwidth]{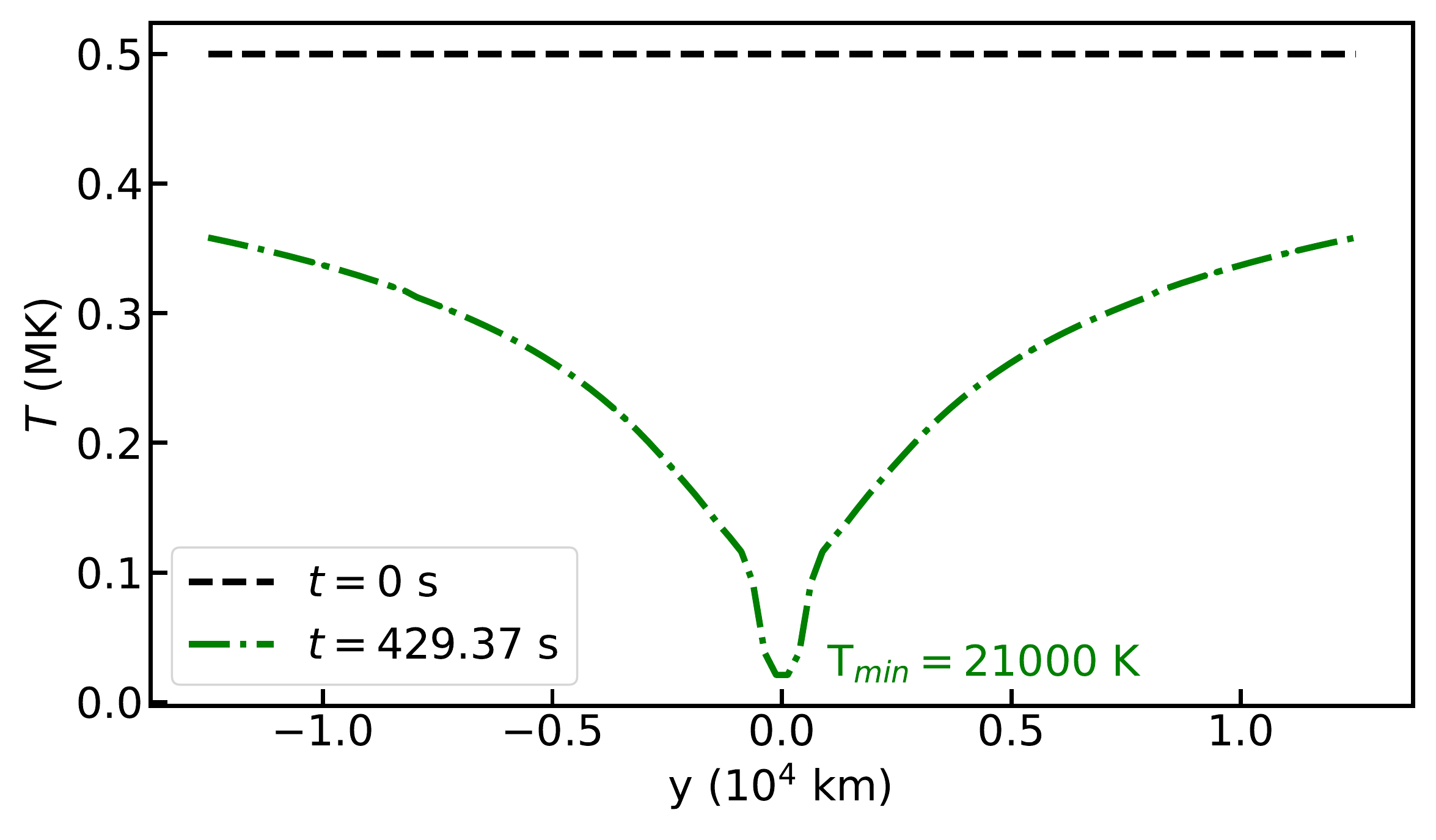}
    \caption{}
    \label{fig:T_ycut}
\end{subfigure}
\caption{Distribution of the (a) radiative cooling (RC) and background heating, $H_{bgr}$, and (b) temperature, $T$ along the vertical cut at $x=0$ between $y=\pm 1.25 \times 10^4$ km for $t=0$ and 429 s. The minimum temperature obtained within this regime at $t=429$ s is $T_{min} = 21000$ K.}
\label{fig:rc_T_ycut}
\end{figure*}

\begin{figure}[hbt!]
    \centering
    \includegraphics[width=0.4\textwidth]{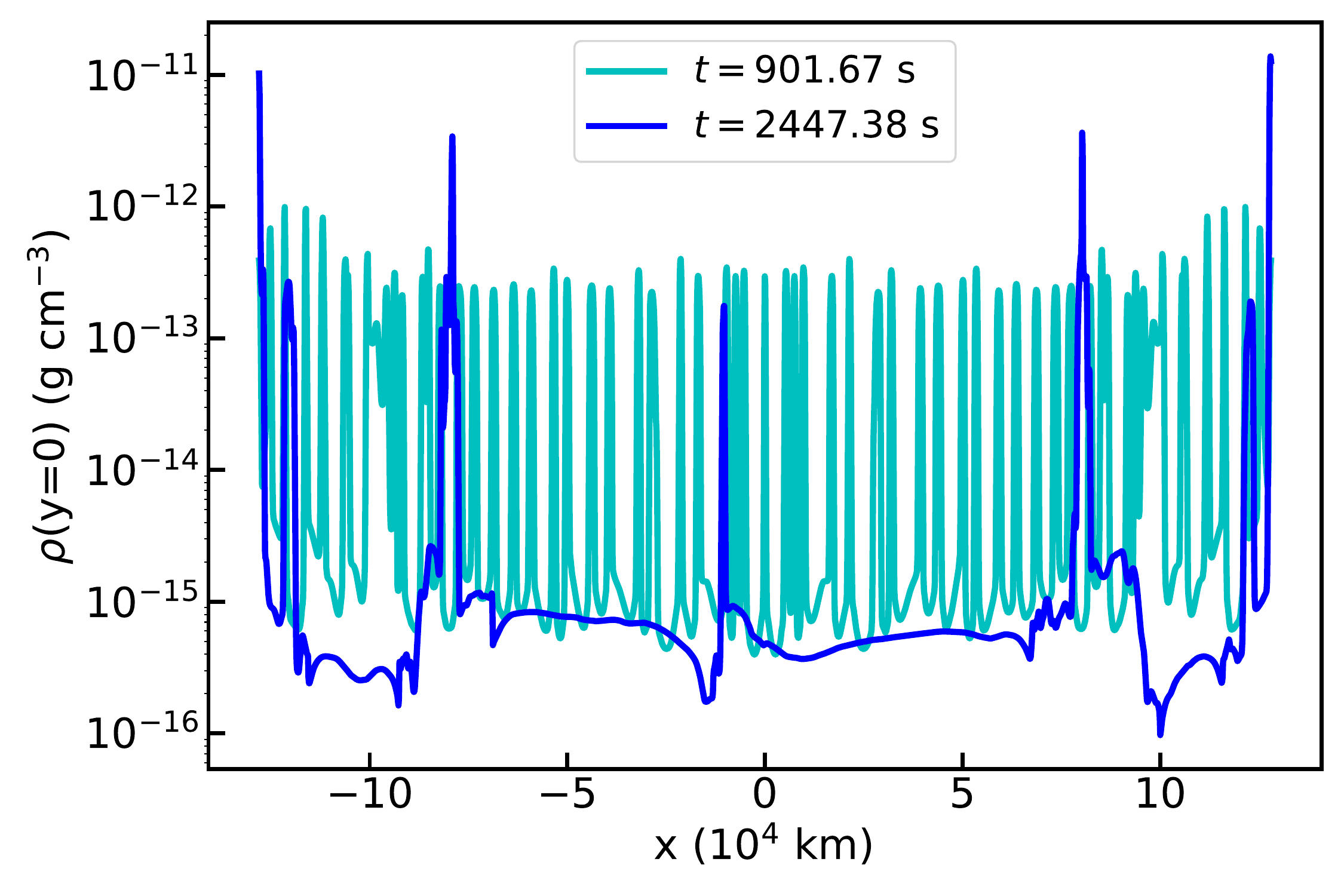}
    \caption{Spatial distribution of plasma density along the $y=0$ cut for two different evolution stages.}
    \label{fig:rho_yo}
\end{figure}
\begin{figure*}[hbt!]
\centering
    \includegraphics[width=0.4\textwidth]{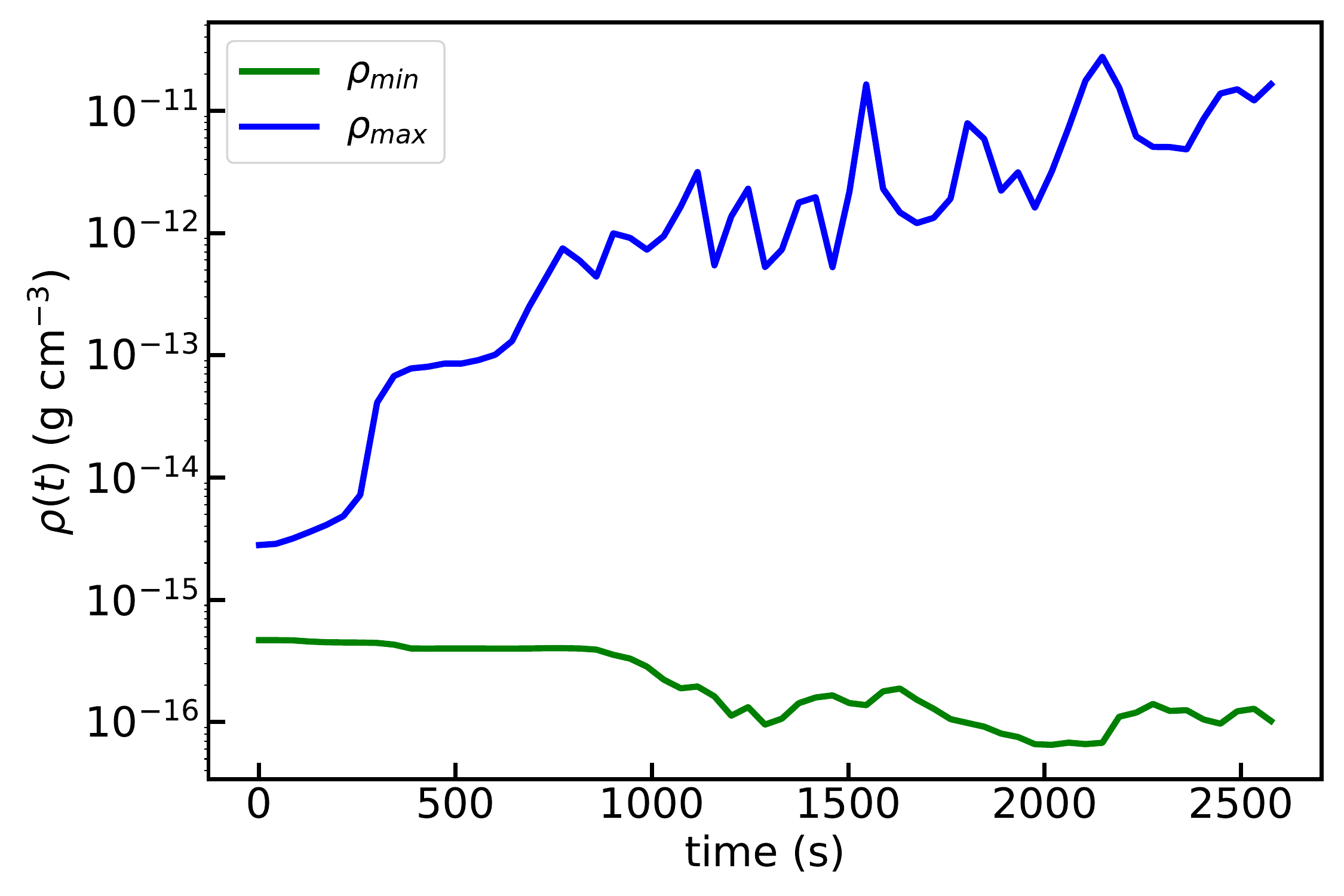}
    \includegraphics[width=0.4\textwidth]{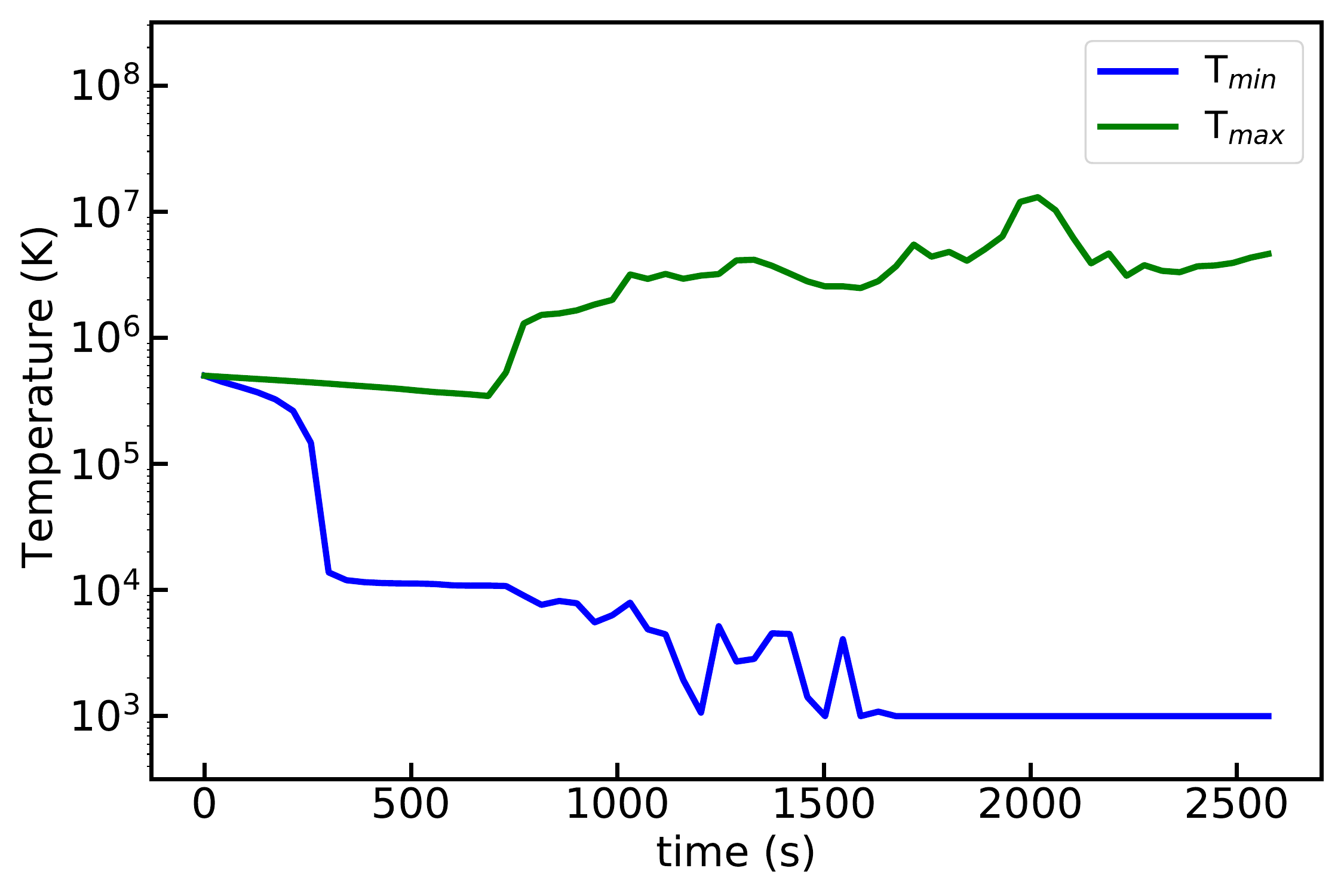}
    \label{fig:rho_T_evol}
\caption{Temporal evolution of the instantaneous maximum and minimum densities (\textit{left}) and temperatures (\textit{right}) within the entire simulation domain.}
\label{fig:rho_T_evol}
\end{figure*}
\begin{figure*}
    \centering
    \includegraphics[scale =0.5]{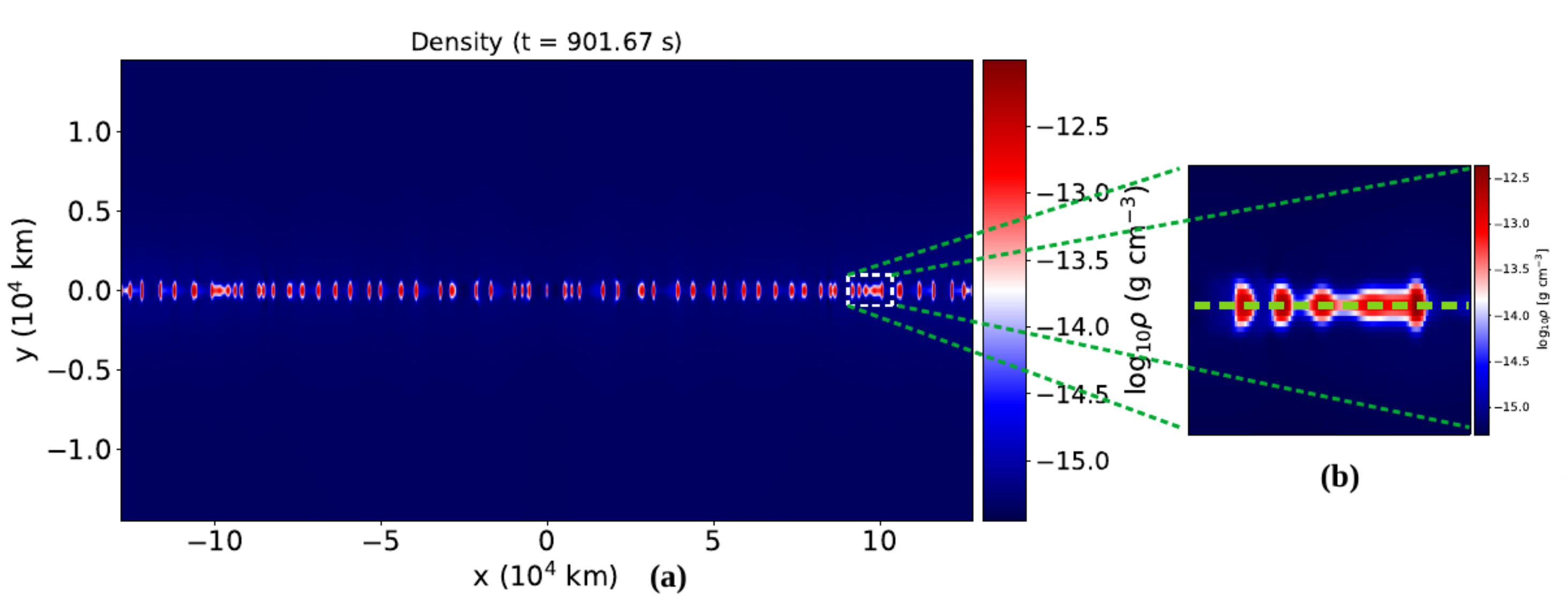}
    \includegraphics[scale =0.5]{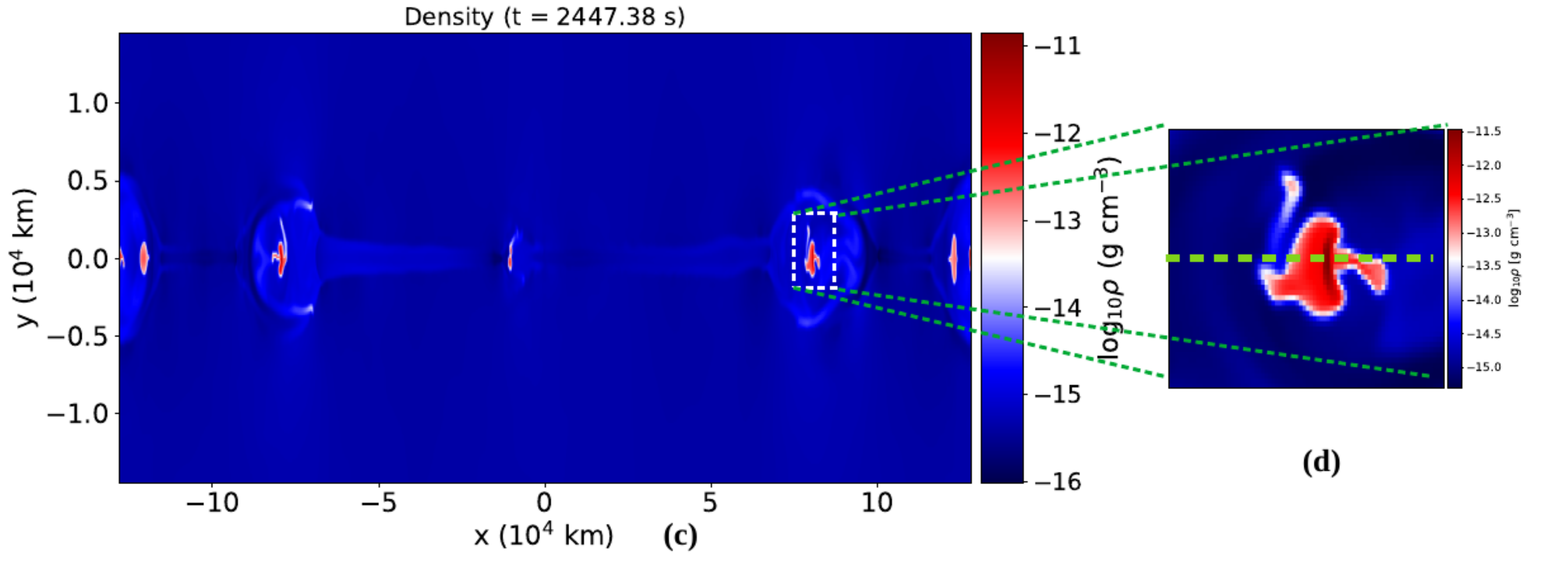}
    \caption{\textit{Top panel:} (a) Distribution of the density for $t=900$ s, within the same domain as Fig. \ref{fig:jzmap}, (b) zoomed version of the selected region. \textit{Bottom panel:} Same as top panel for $t=2447$ s.}
    \label{fig:rho_zoom}
\end{figure*}
\begin{figure*}
    \centering
    \includegraphics[scale =0.5]{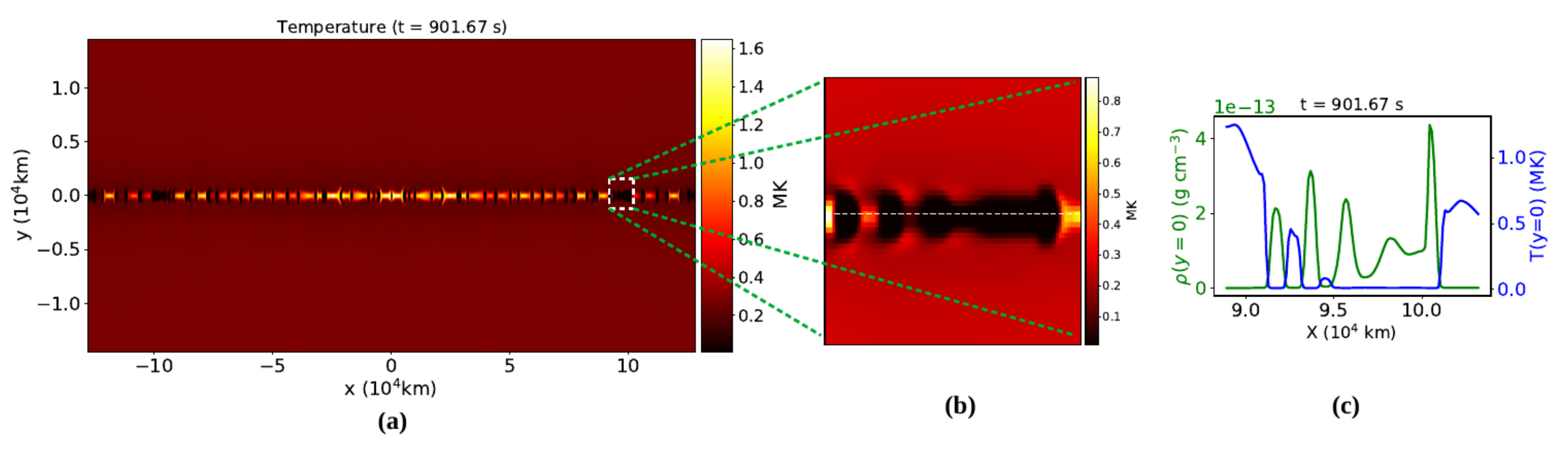}
    \includegraphics[scale =0.5]{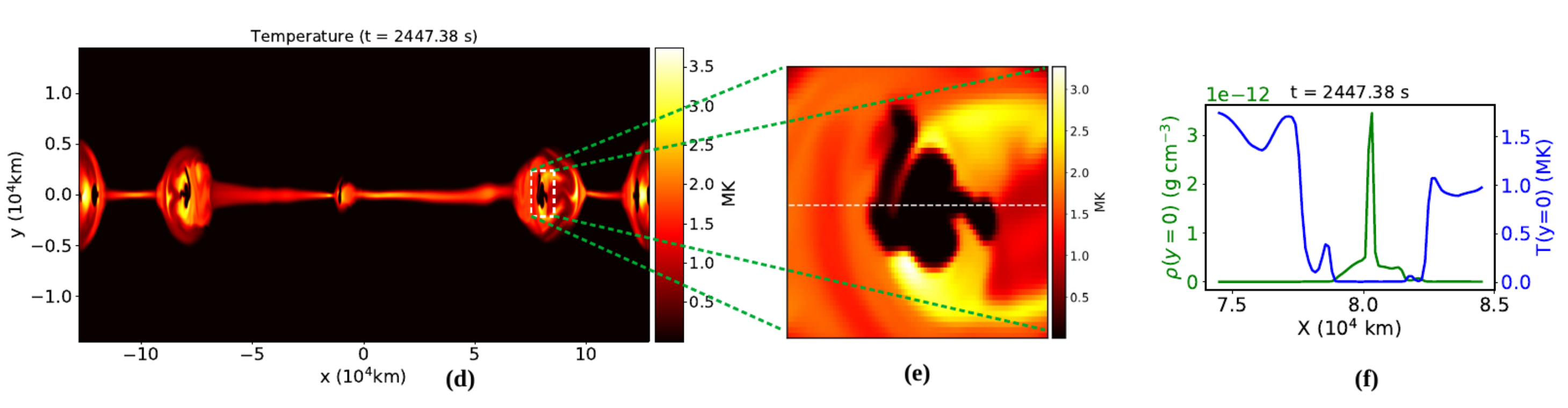}
    \caption{\textit{Top panel:} (a) Distribution of the temperature for $t=900$ s, within the same domain as Fig. \ref{fig:jzmap}, (b) zoomed version of the selected region, (c) Density and temperature distributions along the horizontal cuts marked by the dashed lines in figure \ref{fig:rho_zoom}(b) and \ref{fig:temp_zoom}(b) respectively. \textit{Bottom panel:} Same as top panel for $t=2447$ s.}
    \label{fig:temp_zoom}
\end{figure*}
\begin{figure*}
    \centering
    \includegraphics[scale =0.5]{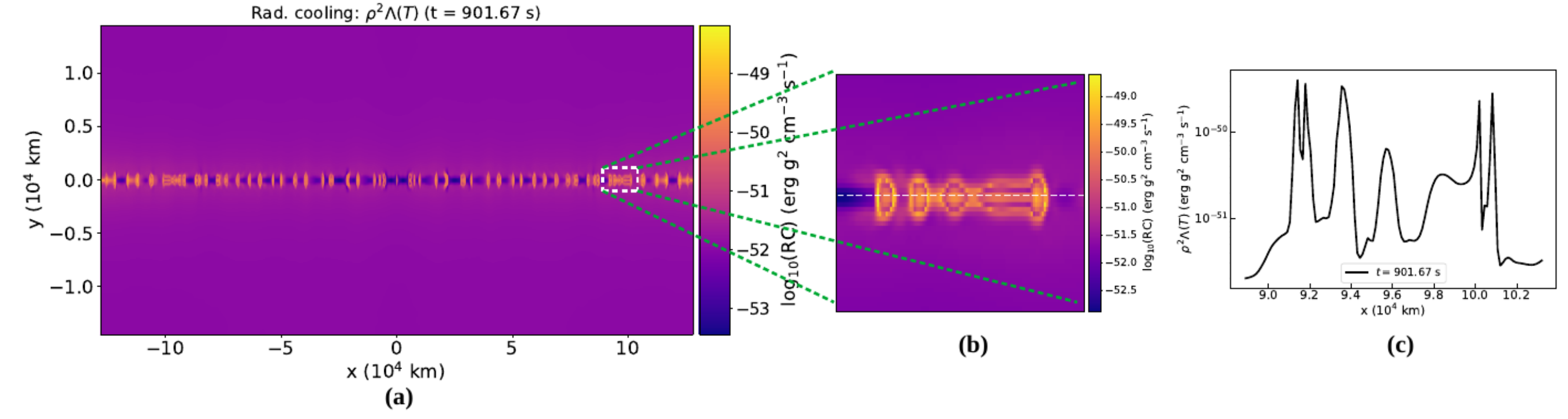}
    \includegraphics[scale =0.5]{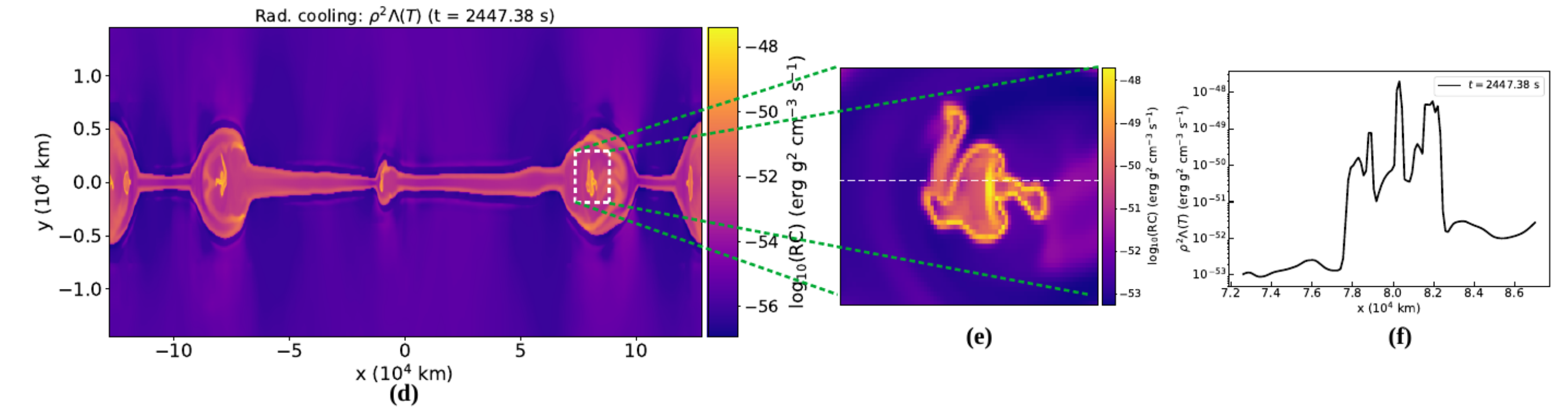}
    \caption{Same as Fig. \ref{fig:temp_zoom} for radiative loss in optically thin medium.}
    \label{fig:rc_zoom}
\end{figure*}

\begin{figure*}[hbt!]
\centering
\begin{subfigure}{0.4\textwidth}
    \includegraphics[width=1.1\textwidth]{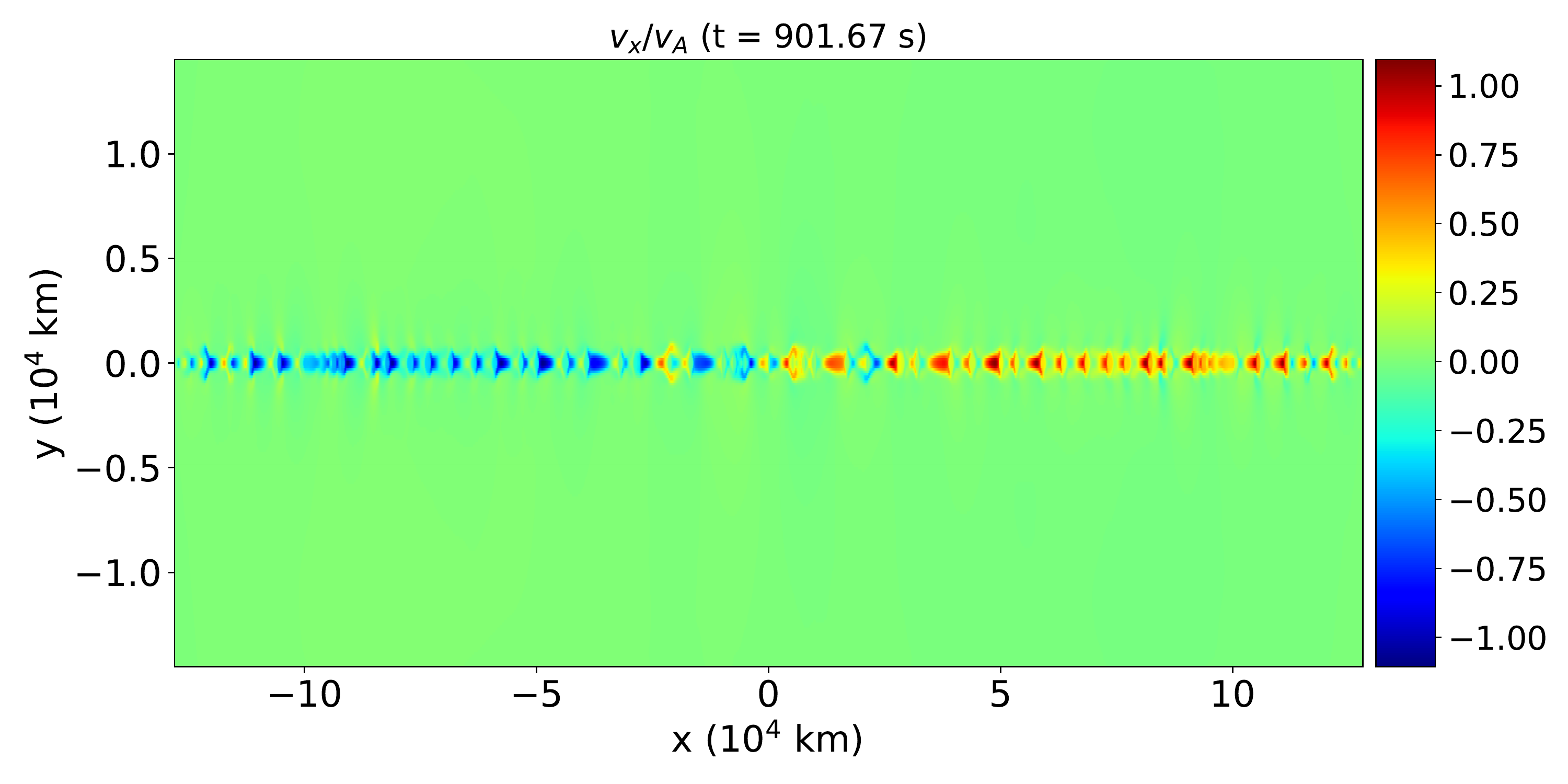}
    \caption{}
    \label{fig:vxmap_t21}
\end{subfigure}
\hspace{1.5 cm}
\begin{subfigure}{0.4\textwidth}
    \includegraphics[width=1.1\textwidth]{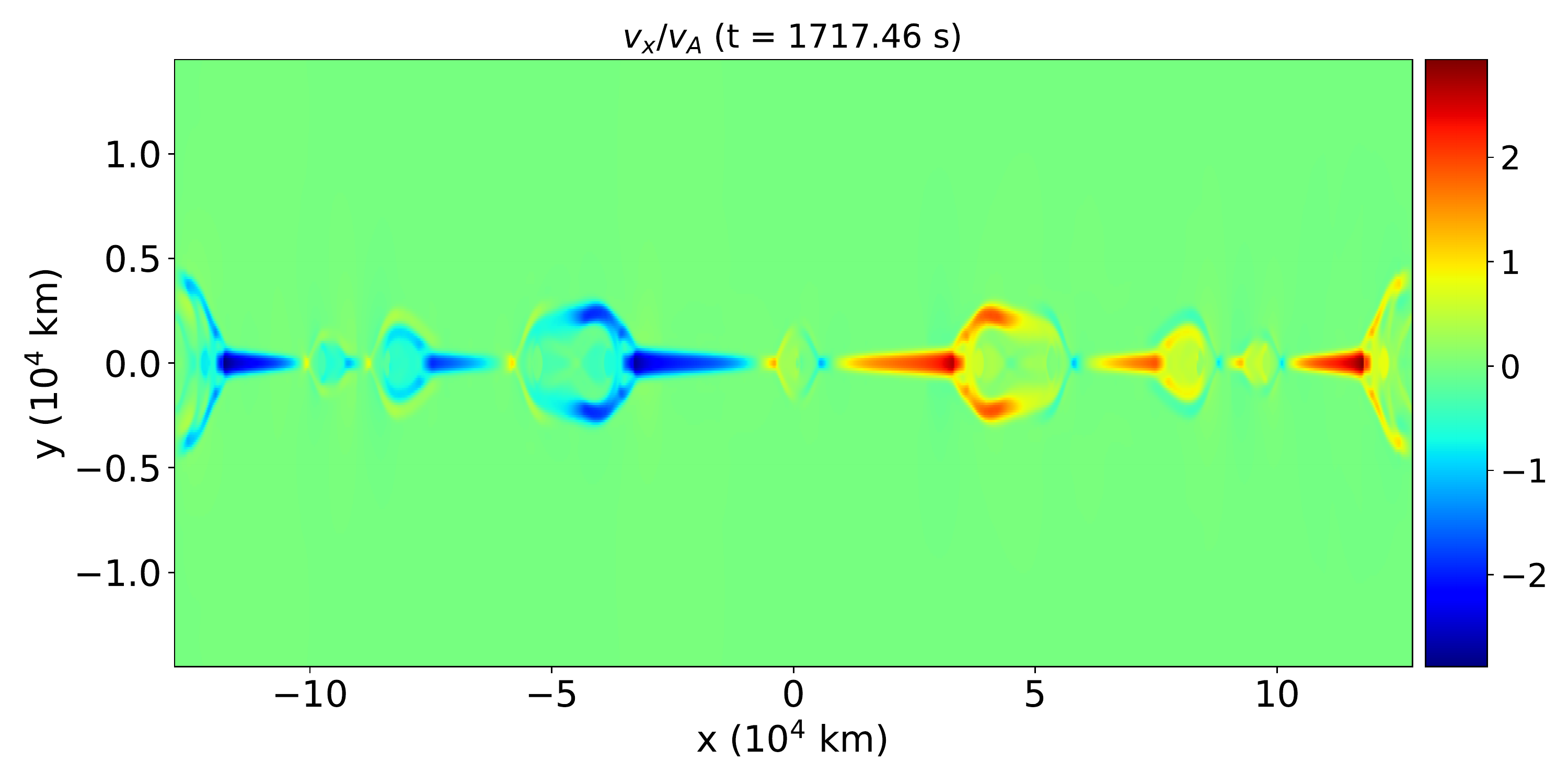}
    \caption{}
    \label{fig:vxmap_t40}
\end{subfigure}
\newline
\begin{subfigure}{0.4\textwidth}
    \includegraphics[width=1.1\textwidth]{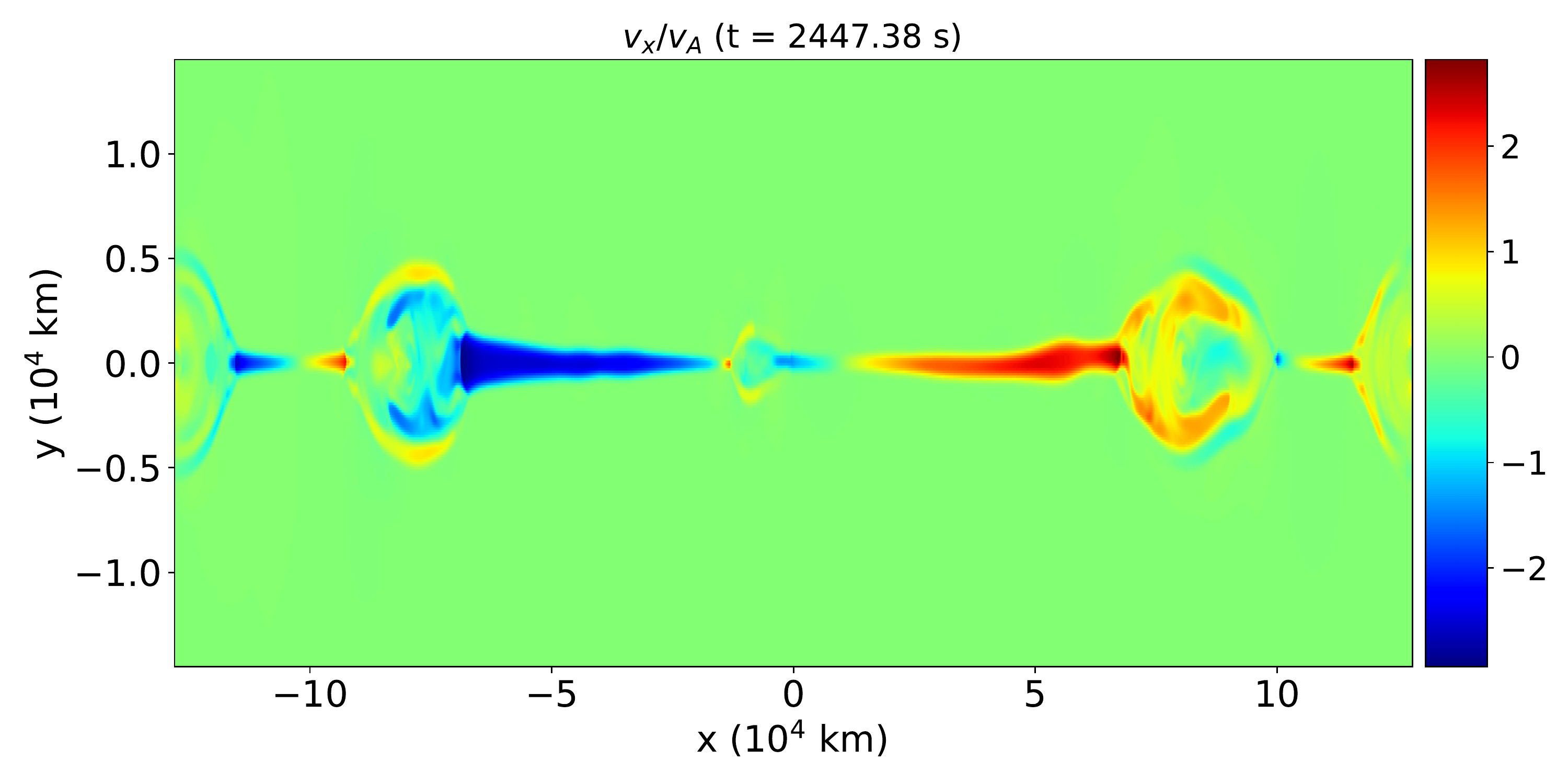}
    \caption{}
    \label{fig:vxmap_t57}
\end{subfigure}
\caption{Distribution of the velocity, $v_x$ within the same domain as Fig. \ref{fig:jzmap}, for different evolution stages. The magnitude of the velocity is scaled with respect to the Alfv{\'e}n velocity, $v_A$. (An animation of the figures is available online).}
\label{fig:vxmap}
\end{figure*}

\begin{figure*}[hbt!]
\centering
\begin{subfigure}{0.4\textwidth}
    \includegraphics[width=1.1\textwidth]{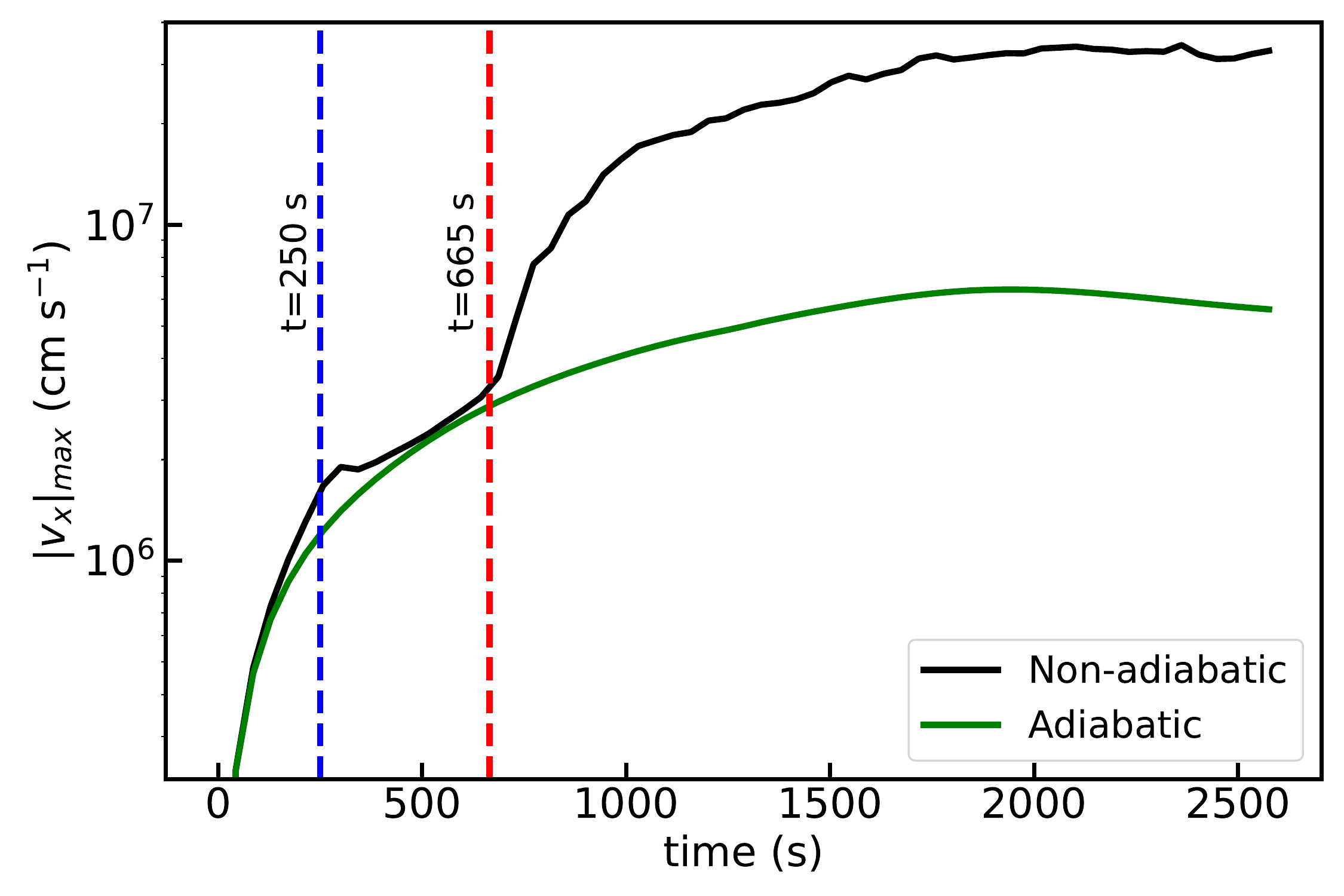}
    \caption{}
    \label{fig:vx_eta001}
\end{subfigure}
\hspace{1 cm}
\begin{subfigure}{0.4\textwidth}
    \includegraphics[width=1.1\textwidth]{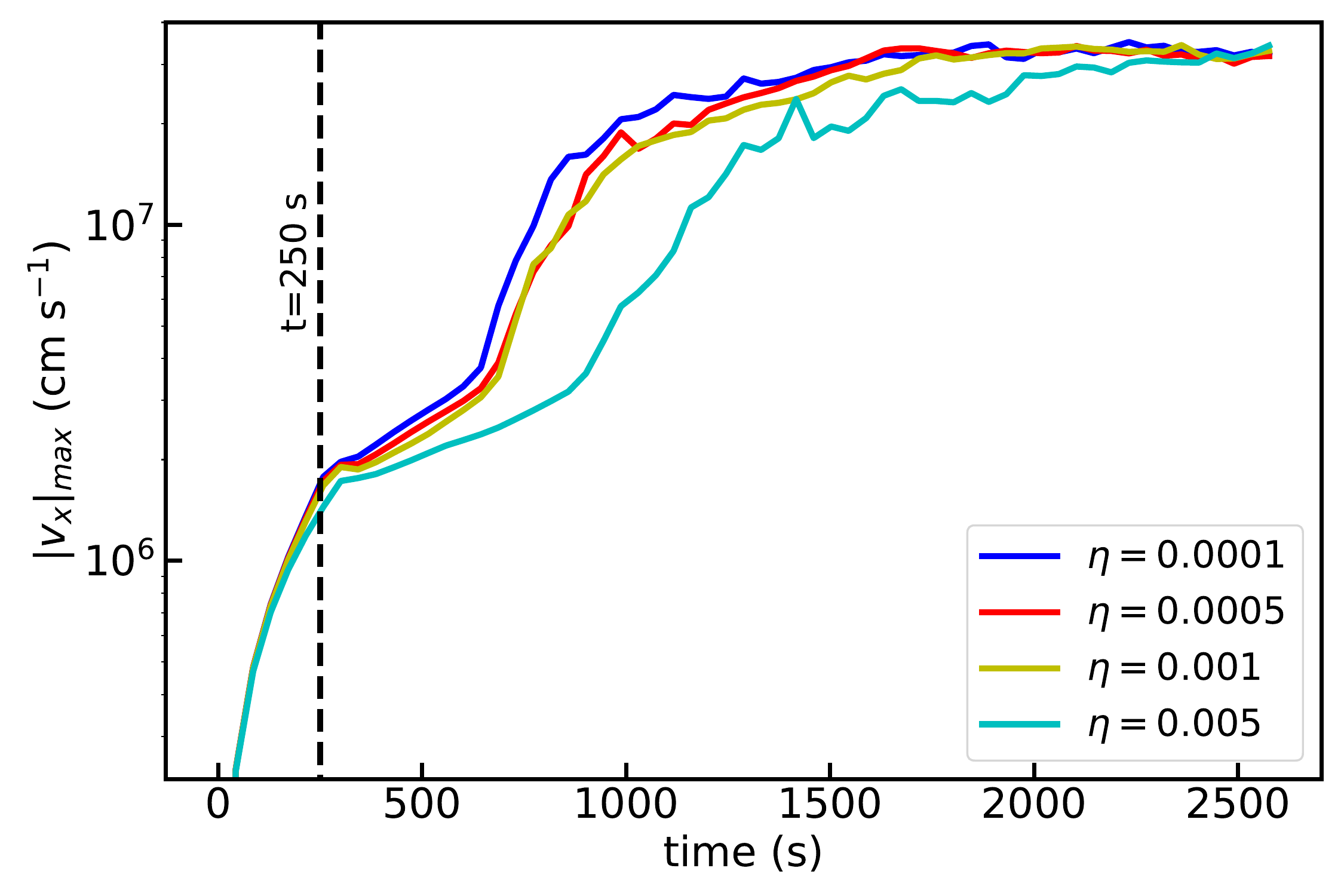}
    \caption{}
    \label{fig:vx_alleta}
\end{subfigure}
\caption{Maximum absolute value of $v_x$ as a function of time. The left panel represents the velocity evolution of the current layer system for two different cases: adiabatic medium, and non-adiabatic medium when radiative energy loss and background heating are incorporated for resistivity, $\eta=0.001$ and plasma-$\beta=0.2$. The right panel is for non-adiabatic evolution for different $\eta$ values keeping all the other parameters the same. The vertical dashed lines represent the different phases of the evolution.}
\label{fig:vx_evol}
\end{figure*}

\begin{figure}[hbt!]
\centering
    \includegraphics[width=0.4\textwidth]{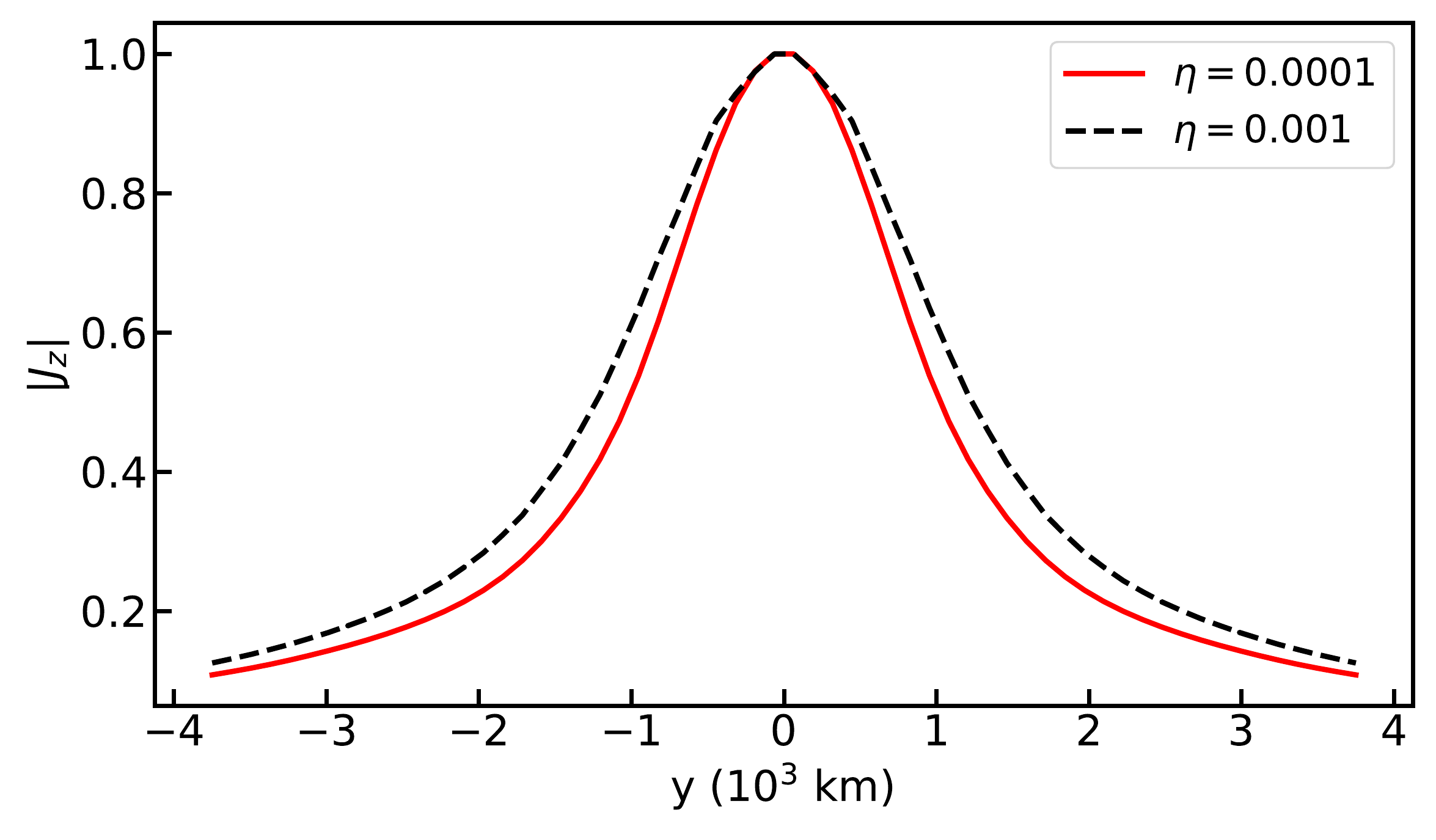}
\caption{Variation of the absolute current density $|J_z|$ (normalized to unity) at $t=214.68$ s, at $x=0$ along the $y$-direction between $\pm 3.75 \times 10^3$ km, for two different resistivities, $\eta = 0.0001$ and 0.001.}
    \label{fig:cs_width}
\end{figure}

\begin{figure}[hbt!]
\centering
\begin{subfigure}{0.35\textwidth}
    \includegraphics[width=1.1\textwidth]{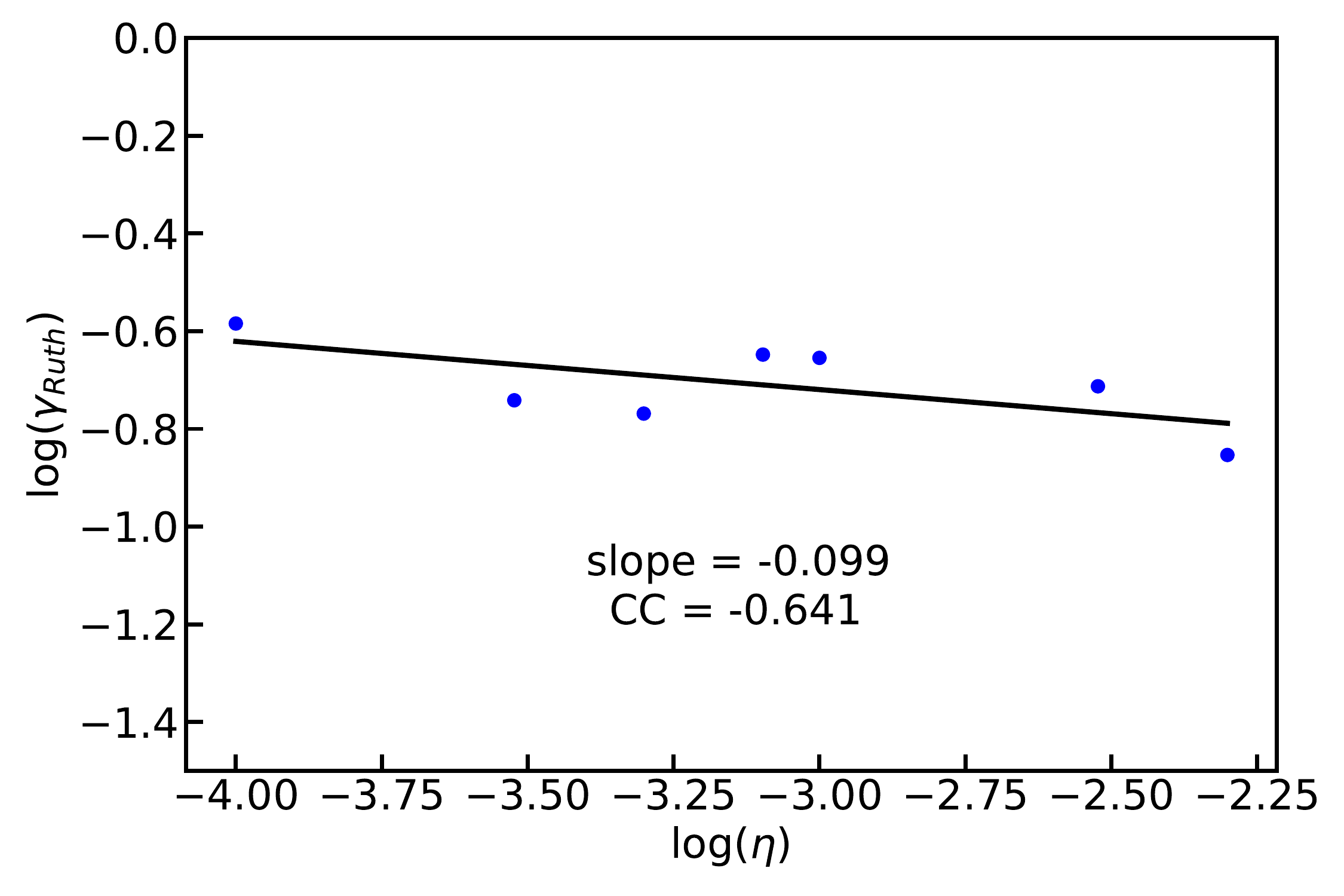}
    \caption{}
    \label{fig:ruth_gr}
\end{subfigure}
\hspace{1 cm}
\begin{subfigure}{0.35\textwidth}
    \includegraphics[width=1.1\textwidth]{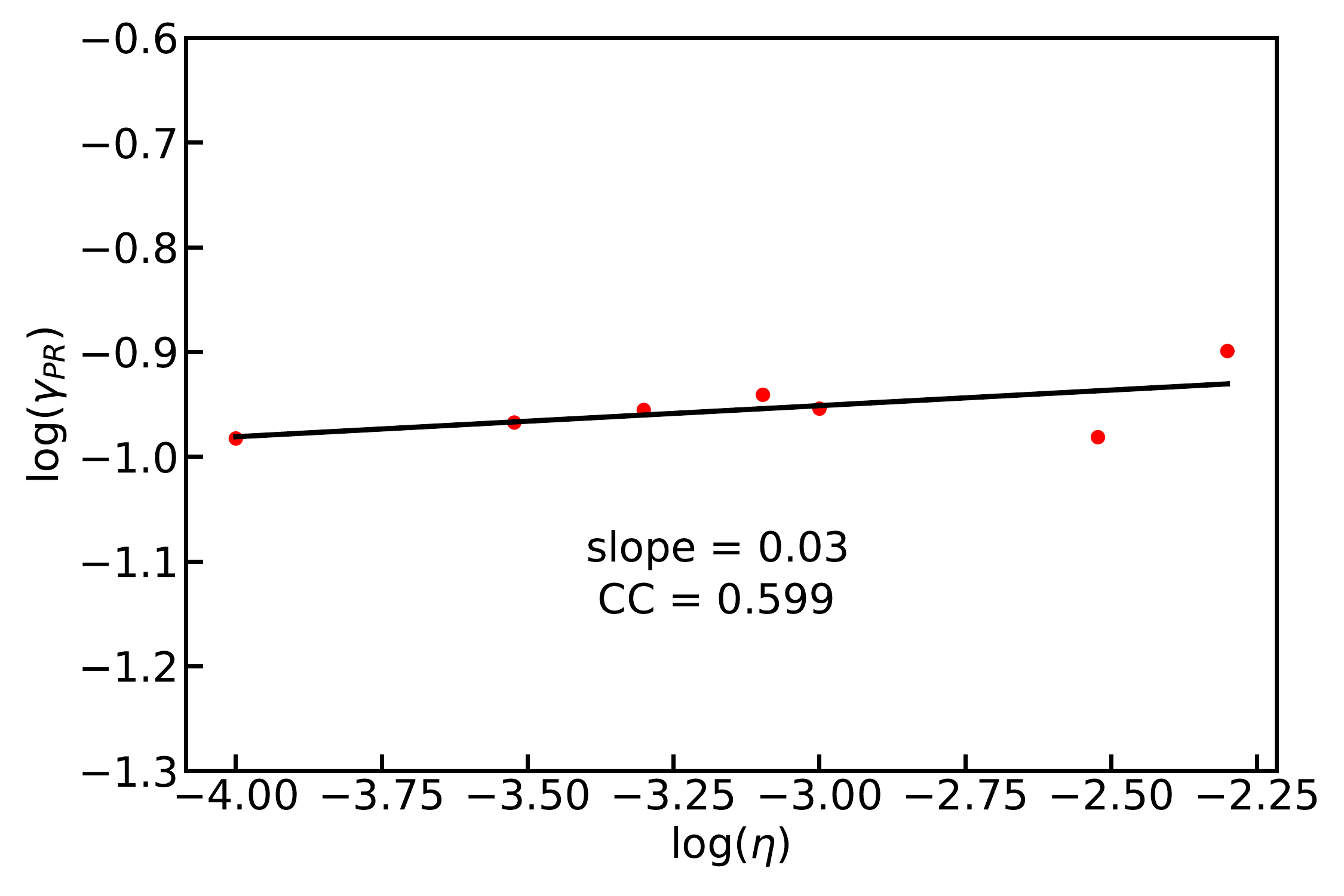}
    \caption{}
    \label{fig:pruth_gr}
\end{subfigure}
\begin{subfigure}{0.35\textwidth}
    \includegraphics[width=1.1\textwidth]{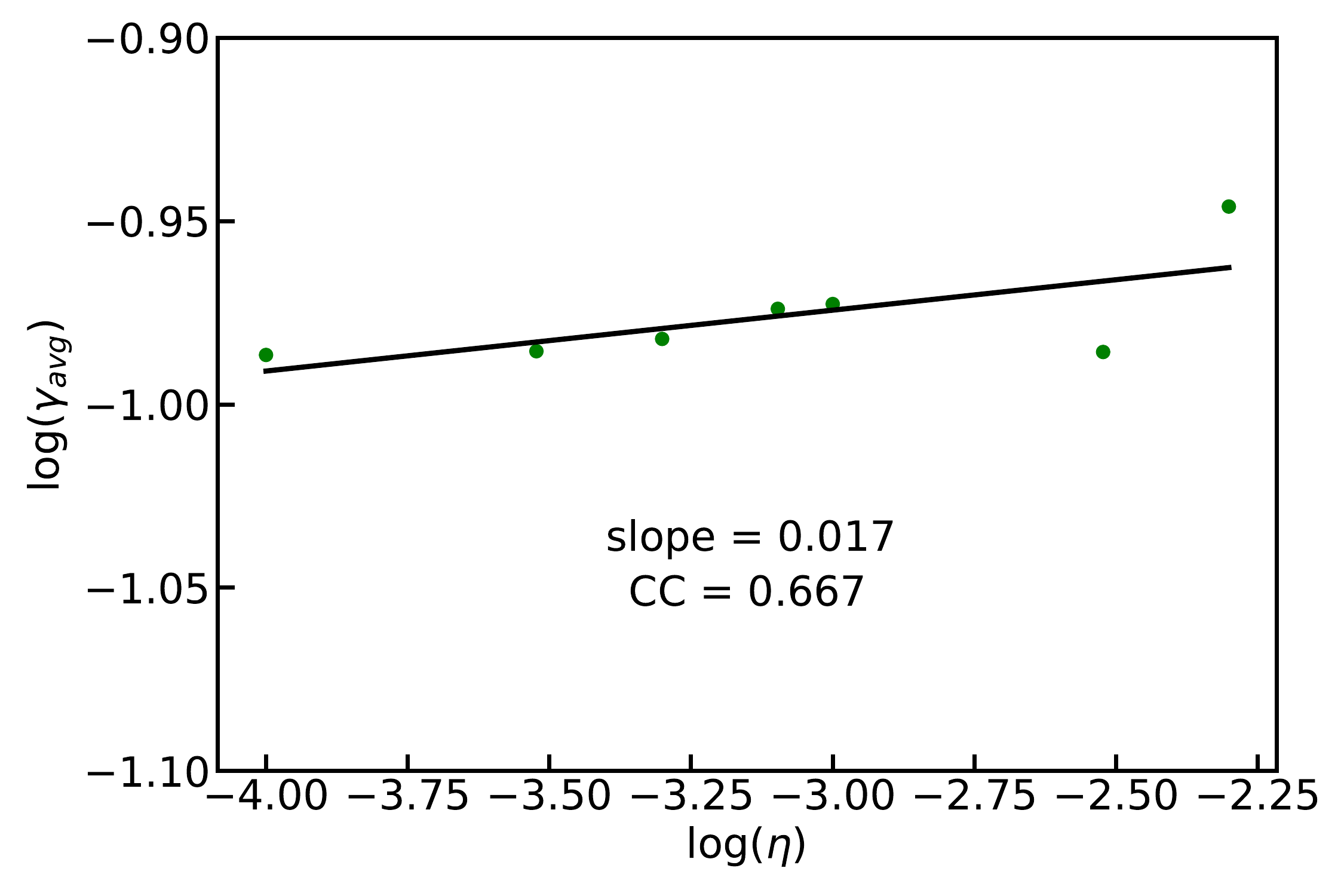}
    \caption{}
    \label{fig:avg_gr}
\end{subfigure}
\caption{Variation of the average growth rates, $\gamma$ with the resistivities for (a) Rutherford, (b) Post-Rutherford, and (c) entire non-linear regions. The growth rates are scaled with respect to the Alfv{\'e}n time scale. The solid lines represent the linear fit of the growth rates vs $\eta$ in the log-log scale. The values of the slope and the correlation coefficients are appended in the corresponding figures.}
\label{fig:gr_eta}
\end{figure}

\begin{figure*}[hbt!]
\centering
\begin{subfigure}{0.4\textwidth}
    \includegraphics[width=1.1\textwidth]{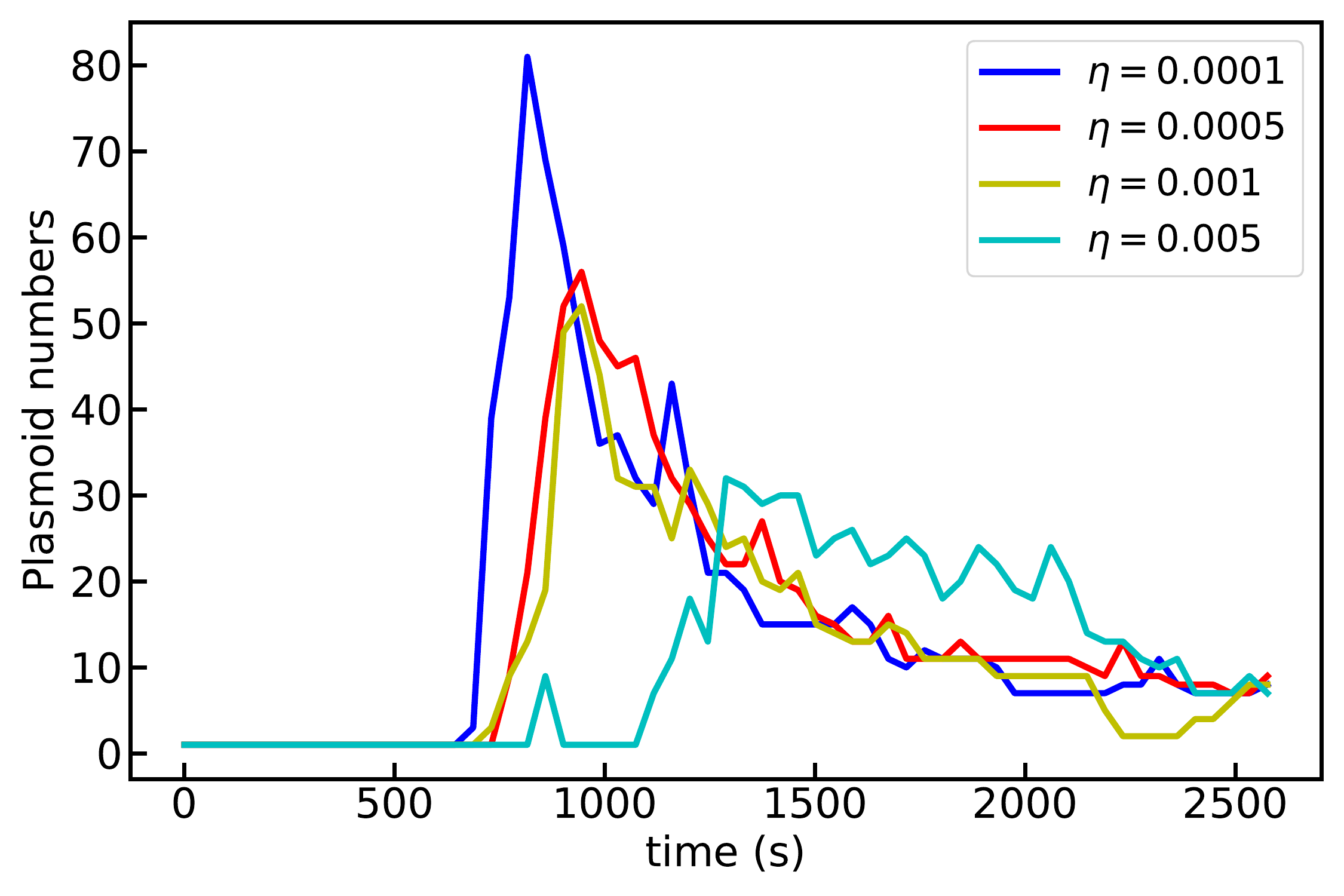}
    \caption{}
    \label{fig:plasmanum_eta}
\end{subfigure}
\hspace{1cm}
\begin{subfigure}{0.4\textwidth}
    \includegraphics[width=1.1\textwidth]{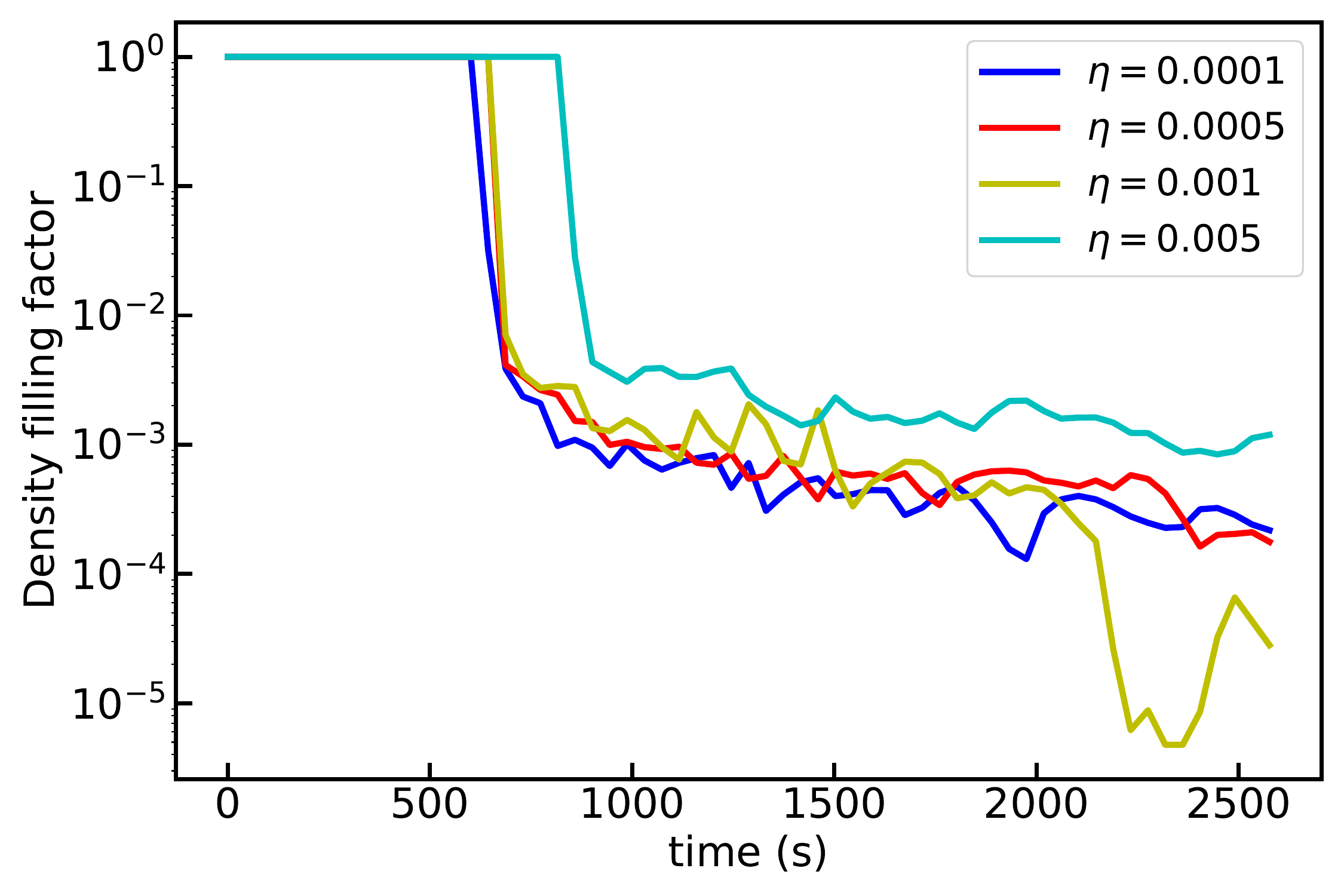}
    \caption{}
    \label{fig:ff_eta}
\end{subfigure}
\caption{Temporal variation of (a) plasmoid numbers and (b) density filling factor for different $\eta$ with plasma-$\beta=0.2$.}
\label{fig:plasma_ff_eta}
\end{figure*}

\begin{figure}[hbt!]
\centering
\includegraphics[width=0.4\textwidth]{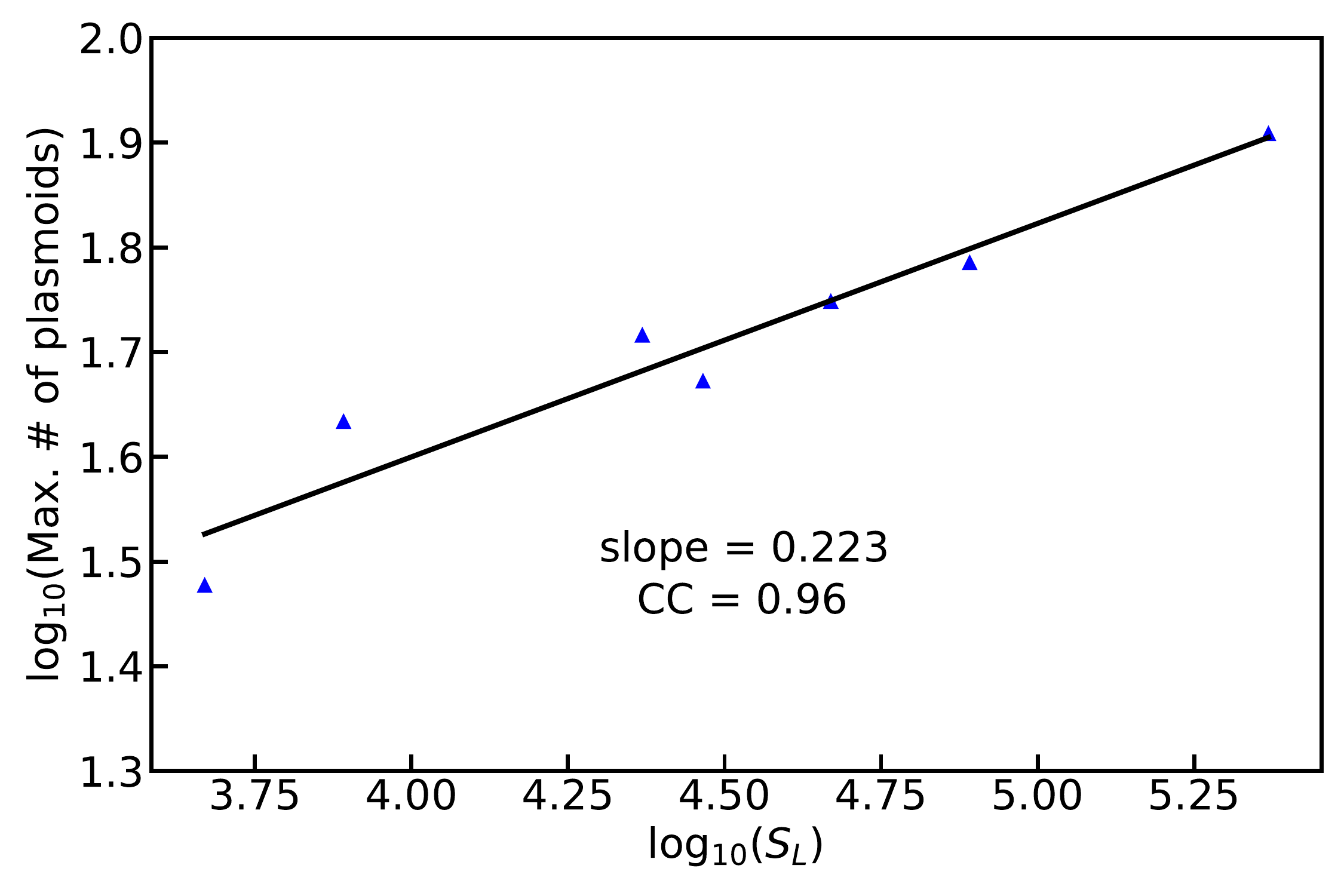}
\caption{Maximum number of the plasmoids vs Lundquist number, $S_L$ obtained from different simulation runs (for plasma-$\beta$=0.2) is shown by the blue triangles. The solid line represents the linear fit in the log-log scale of the plasmoid numbers vs $S_L$. The slope and Pearson's correlation coefficient for the linear fit are 0.223 and 0.96 respectively.}
\label{fig:plasma_num_sl}
\end{figure}

\begin{figure*}[hbt!]
\centering
\begin{subfigure}{0.4\textwidth}
    \includegraphics[width=1.1\textwidth]{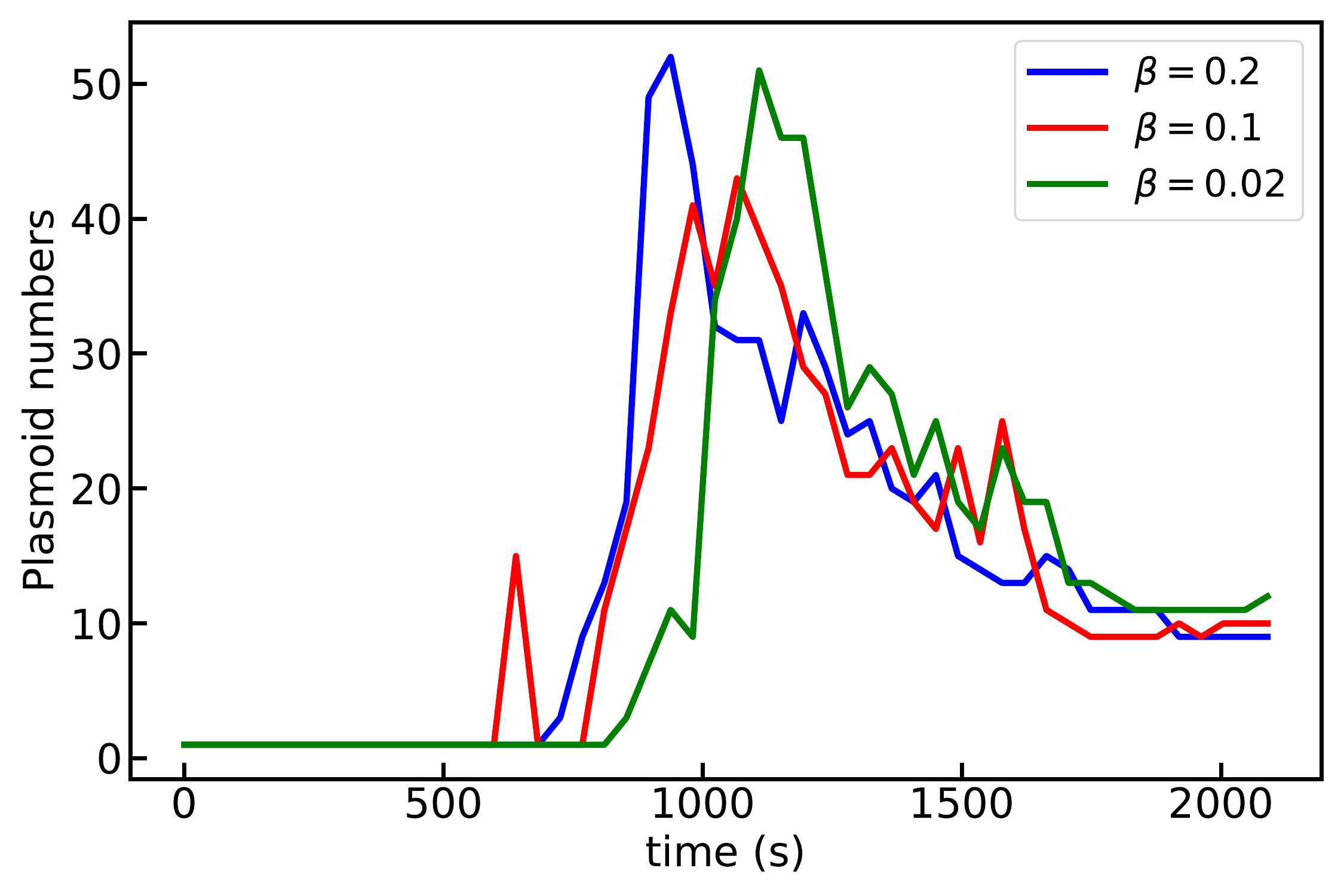}
    \caption{}
    \label{fig:plasmanum_beta}
\end{subfigure}
\hspace{1 cm}
\begin{subfigure}{0.4\textwidth}
    \includegraphics[width=1.1\textwidth]{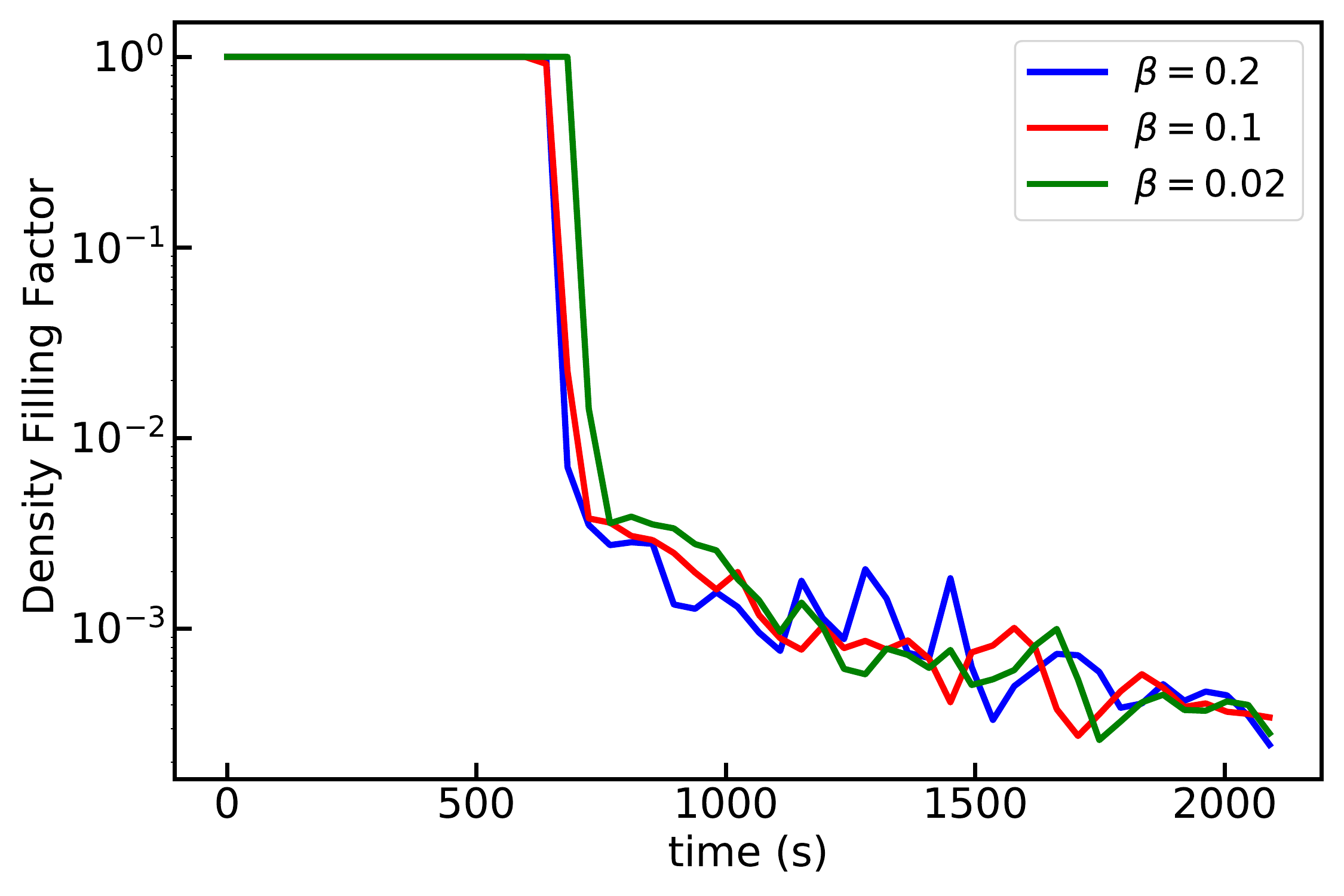}
    \caption{}
    \label{fig:ff_beta}
\end{subfigure}
\caption{Temporal variation of (a) plasmoid numbers, and (b) density filling factor for different plasma-$\beta$ with $\eta=0.001$.}
\label{fig:plasma_ff_beta}
\end{figure*}
\begin{figure*}[hbt!]
\centering
\begin{subfigure}{0.4\textwidth}
    \includegraphics[width=1.1\textwidth]{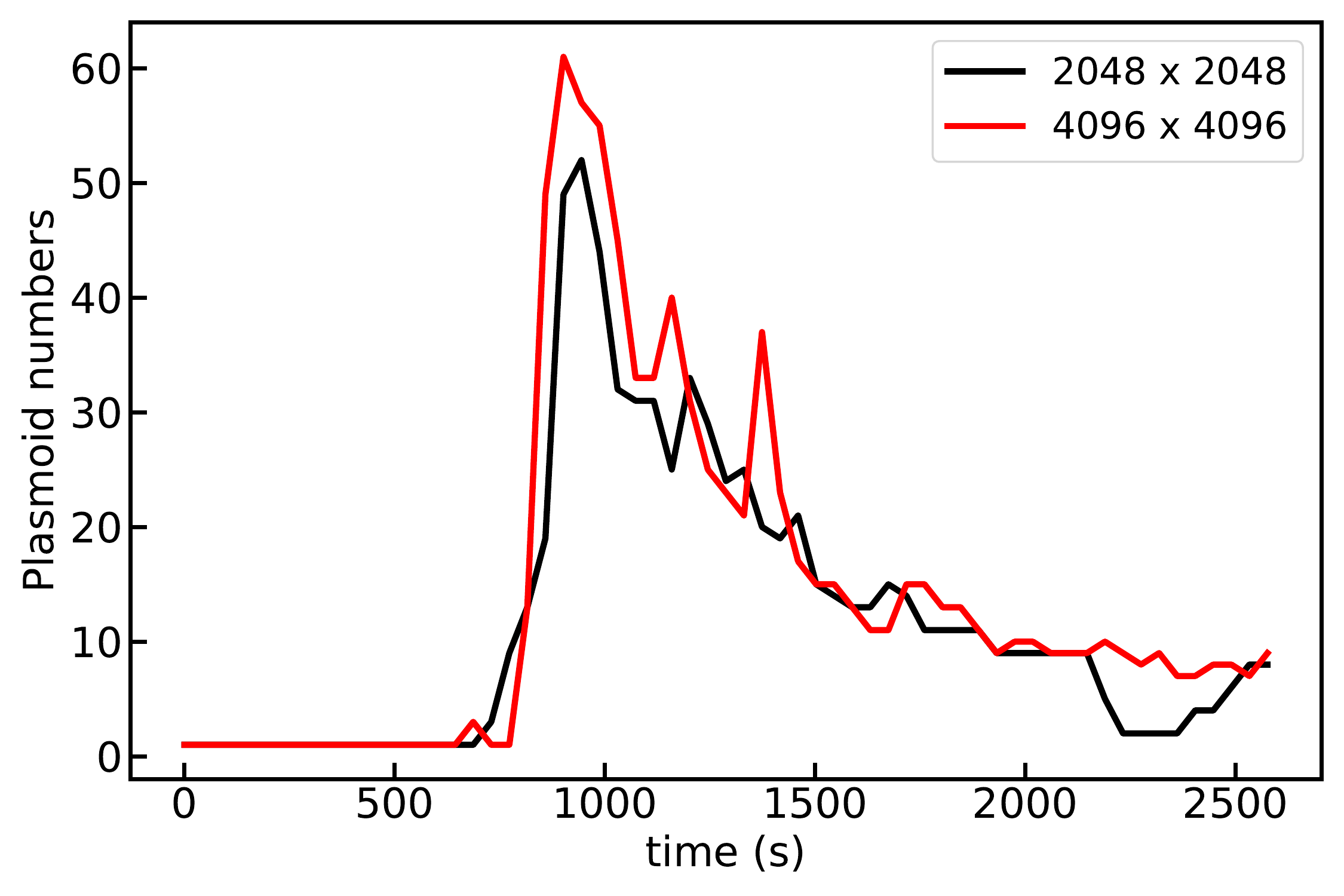}
    \caption{}
    \label{fig:plasmanum_reso}
\end{subfigure}
\hspace{1 cm}
\begin{subfigure}{0.4\textwidth}
    \includegraphics[width=1.1\textwidth]{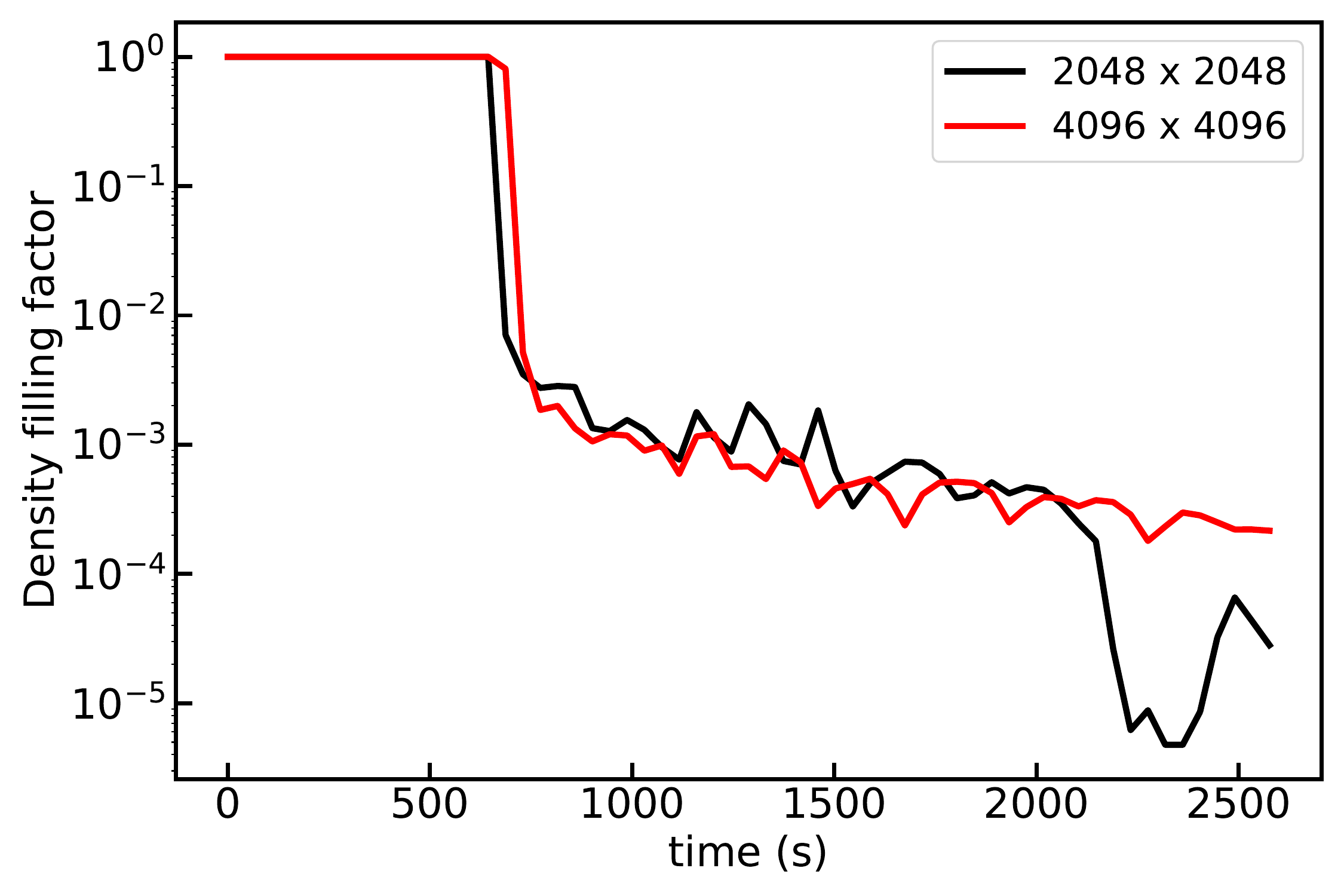}
    \caption{}
    \label{fig:ff_reso}
\end{subfigure}
\caption{Temporal variation of (a) plasmoid numbers, and (b) density filling factor for different numerical resolution for $\eta=0.001$ and plasma-$\beta=0.2$.}
\label{fig:plasma_ff_reso}
\end{figure*}

\begin{figure*}[hbt!]
\centering
\begin{subfigure}{0.4\textwidth}
    \includegraphics[width=1.1\textwidth]{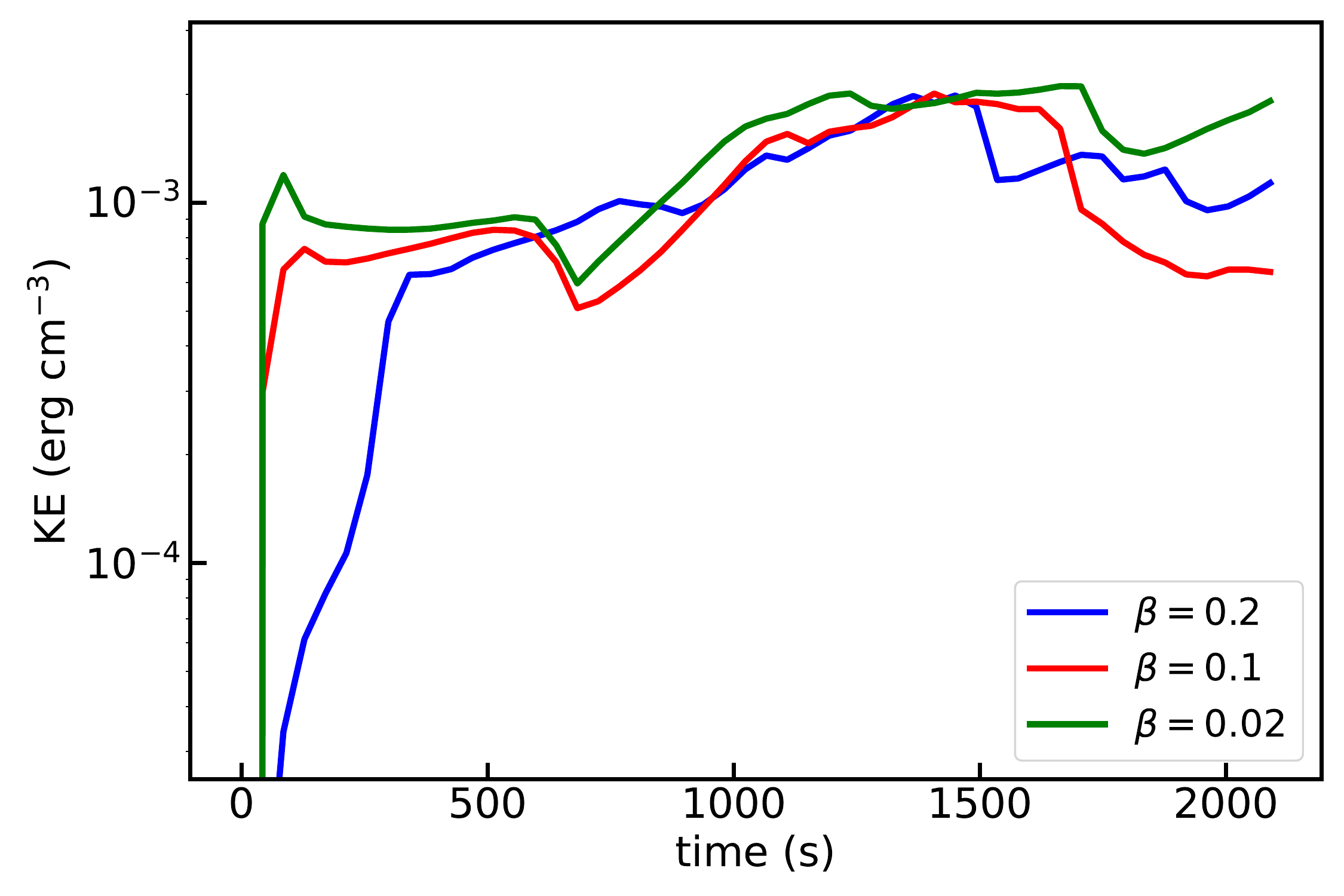}
    \caption{}
    \label{fig:ke}
\end{subfigure}
\hspace{1 cm}
\begin{subfigure}{0.4\textwidth}
    \includegraphics[width=1.1\textwidth]{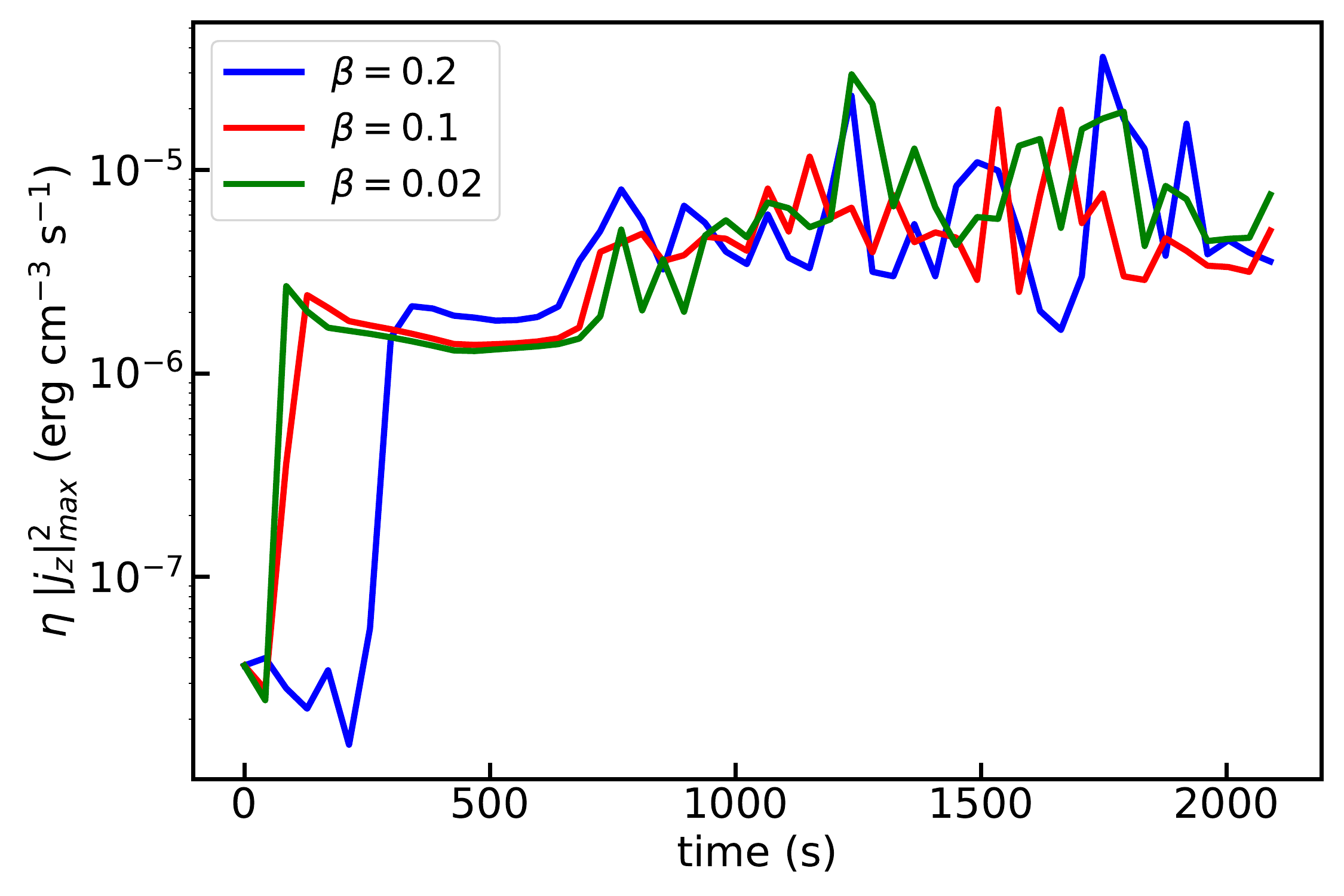}
    \caption{}
    \label{fig:ohm_heat}    
\end{subfigure}
\caption{Temporal variation of (a) kinetic energy density, and (b) ohmic heating rate for $\eta=0.001$ and different plasma-$\beta$.}
\label{fig:heating}
\end{figure*}


\section{Numerical setup of the model}\label{setup}

For understanding the evolution of the thermal instability in association with the tearing mode, we use a resistive 2D MHD simulation using MPI-parallelised Adaptive Mesh Refinement Versatile Advection Code (MPI-AMRVAC)\footnote{open source at: \url{http://amrvac.org}} \citep{2012JCoPh.231..718K, 2014ApJS..214....4P, 2018ApJS..234...30X, keppens2021}. 
The spatial domain of the simulation setup is $-12.8 \times 10^4$ km to $12.8 \times 10^4$ km along both $x$ and $y$ directions with maximum five levels of adaptive mesh refinement (AMR) between $-2 \times 10^4\ \mathrm {km} \leq y \leq 2 \times 10^4\ \mathrm {km}$ and $-12.8 \times 10^4\ \mathrm {km} \leq x \leq 12.8 \times 10^4\ \mathrm {km}$, effectively achieving a maximum spatial resolution of 2048 $\times$ 2048, which makes the smallest cell of size 125 km. The triggering of the (de)refinement is based on the errors estimated by the density (gradient) of an instantaneous time step. To explore the non-adiabatic effects in the evolution of the tearing instability in the resistive MHD regime, we solve the following normalised MHD equations numerically,

\begin{align} \label{eq:mhd1}
   & \frac{\partial \rho}{\partial t} + \nabla \cdot ({\bf v}\rho) = 0, \\ \label{eq:mhd2}
   & \frac{\partial (\rho {\bf v})}{\partial t}+ \nabla \cdot (\rho {\bf v} {\bf v}+p_{tot}{\bf I}-{\bf B}{\bf B})=0,\\ \label{eq:mhd3}
   & \frac{\partial \mathcal{E}}{\partial t} + \nabla \cdot (\mathcal{E}{\bf v}+p_{tot}{\bf v}-{\bf B}{\bf B}\cdot {\bf v})= \eta \textbf{J}^2 - {\bf B} \cdot \nabla \times (\eta {\bf J})\\\nonumber 
   & - \rho^2 \Lambda(T)+H_{bgr},\\ \label{eq:mhd4}
   & \frac{\partial {\bf B}}{\partial t} + \nabla \cdot ({\bf v}{\bf B}-{\bf B}{\bf v}) + \nabla \times (\eta {\bf J}) = 0\,,\\ \label{eq:mhd5}
   & \nabla \cdot \textbf{B}=0 \,,\\ \label{eq:mhd6}
   & \textbf{J}=\nabla \times \textbf{B} \,.
\end{align}
Here, $\textbf{I}$ is the unit tensor, and the quantities $\rho, T, {\bf B}$, ${\bf v}$, and $\eta$ have their usual meaning of the mass density, temperature, magnetic field vector, velocity, and resistivity respectively. The total pressure, $p_{tot}$ adds the plasma and the magnetic pressure 
\begin{align}\label{eq:ptot}
    p_{tot} = p+\frac{B^2}{2},
\end{align}
where, the thermodynamic quantities are linked through the ideal gas law: $\displaystyle{p={\rho k_B T}/{\mu m_H}}$, where $p$ is the plasma pressure, $k_B$ is the Boltzman constant, $\mu=0.6$ is the coronal abundance due to the fully ionized plasma of H and He atoms with the abundance ratio of $10:1$, and $m_H$ is the proton mass. The total energy density is given by
\begin{align}\label{eq:energy_density}
    \mathcal{E}=\frac{p}{\gamma_{gc} -1} + \frac{\rho v^2}{2} + \frac{B^2}{2},
\end{align}
where, $\gamma_{gc} = 5/3$ is the ratio of specific heats for the monoatomic gas. 
The solenoidal condition of magnetic field, and the current density, $\textbf{J}$ are given by the equations (\ref{eq:mhd5}) and (\ref{eq:mhd6}) respectively. The non-adiabatic effect due to the radiative cooling of the optically thin medium, which is relevant for the solar corona is incorporated by the third term in the RHS of equation (\ref{eq:mhd3}). The optically thin cooling due to the radiation depends on the local density, and the temperature-sensitive cooling model $\Lambda(T)$. In this work, we have used the combined cooling model developed by \cite{2008ApJ...689..585C} and \cite{1972ARA&A..10..375D}, which we call as `Colgan\_DM' model, shown in Fig. \ref{fig:clogan_dm}. The details of different radiative cooling curves and their effects on the formation of condensations are reported in \cite{2021A&A...655A..36H}. In order to maintain a thermal equilibrium in the initial state, we use the background heating, $H_{bgr}$ (last term in the RHS of equation (\ref{eq:mhd3})) in such a way that it compensates the radiative loss at the initial state. Hence, we take the background heating as
\begin{align}\label{eq:Hbr}
    H_{bgr} = \rho_i^2 \Lambda(T_i),
\end{align}
where, $\rho_i$ and $T_i$ are the equilibrium density and temperature respectively, and therefore the background heating is constant with time (but it varies in space as explained in the following section). Note that for simplicity, we here study an idealized current sheet setup, where we ignore gravity and the role of (anisotropic) thermal conduction.\\

To study the long term behaviour of the evolution of a current layer configuration subjected to the resistive MHD and non-adiabatic effects of radiative cooling, we set a 2D square domain in the $x-y$ plane that spans between -12.8 to 12.8 (in dimensionless unit) along $x$ and $y$ directions. The unit density, temperature and  length scale which serve to normalize the simulation are set as $\bar{\rho}=2.34 \times 10^{-15}$ g cm$^{-3}$, $\bar{T}=10^6$ K and $\bar{L}=10^9$ cm respectively, which are typical values for the solar corona. The magnetic field in the initial setup is taken as a non-force free planar field given by,
\begin{align}\label{eq:Bx}
    & B_x = B_0\ \mathrm{tanh}(y/l_s)\,,\\\label{eq:By}
    & B_y = 0.
\end{align}
This implies that the $B_x$ component realizes a polarity inversion around $y=0$, and hence a current sheet is formed at $y=0$ that spans between $x=[-12.8,12.8]\times \bar{L}$. We set the background field amplitude, $B_0=1$ (which corresponds to 2 G in physical unit), and $l_s=0.5$, which sets the total width of the current sheet to $2l_s$. The planar magnetic field $B_x(y)$ is further perturbed to trigger some tearing-type evolution, as follows
\begin{align}\label{eq:delbx}
   & \delta B_x = -\frac{2\pi \psi_0}{l_y}\ \mathrm{cos}\bigg(\frac{2\pi x}{l_x}\bigg)\ \mathrm{sin}\bigg(\frac{2\pi y}{l_y}\bigg),\\ \label{eq:delby}
   & \delta B_y = + \frac{2\pi \psi_0}{l_x}\ \mathrm{sin}\bigg(\frac{2\pi x}{l_x}\bigg)\ \mathrm{cos}\bigg(\frac{2\pi y}{l_y}\bigg),
\end{align}
where the geometric parameters $l_x = l_y = 25.6 \times \bar{L}$ match the domain sizes of the simulation, and the perturbation amplitude $\psi_0=0.1$ is 10\% of the magnetic field amplitude $B_0$. It is to be noted that equations (\ref{eq:delbx}) and (\ref{eq:delby}) satisfy the condition $\nabla \cdot \delta {\bf B}=0$. The initial density profile is taken as
\begin{align}\label{eq:rho_in}
    \rho_i = \rho_0 + \mathrm{cosh}^{-2}(y/l_s),
\end{align}
where $\rho_0=0.2$ (which corresponds to $4.68 \times 10^{-16}$ g cm$^{-3}$ in physical unit) is the density outside the current sheet. The initial density profile has the peak value of $2.808 \times 10^{-15}$ g cm$^{-3}$ at $y=0$ and gradually converges to $\rho_0$ for $|y|>0$. The initial equilibrium temperature $T_i$ is taken as a constant value of 0.5 MK throughout the simulation domain in order to fix the plasma-$\beta = 0.2$, which is below unity as appropriate for the solar corona. The initial variation in density, together with the uniform temperature, realizes a pressure variation $p(y)$ that exactly balances the Lorentz force associated with the field $B_x(y)$. This implies that in a simulation where no magnetic field perturbation is applied, and where the resistivity is set to zero (ideal MHD), we actually have a force-balanced and thermally balanced environment. In our simulations below, resistive effects will modify the temperature and hence drive the system away from the thermal equilibrium balance between losses and heating $H_{bgr}$. Note that the finite resistivity is crucial to allow tearing. If we simulate without non-adiabatic effects included, and just evolve the system in resistive MHD at the used constant and uniform resistivity values, we get an evolution towards a standard reconnection experiment with a central Sweet-Parker type current sheet in between a growing island structure (at both periodic sides).

After this initial setup the system evolves as governed by the equations (\ref{eq:mhd1}-\ref{eq:mhd6}). This set of equations is solved using a three-step Runge-Kutta time integration with a third-order slope limited reconstruction method \citep{2009JCoPh.228.4118C}, and Harten-Lax-van Leer (HLL) flux scheme \citep{doi:10.1137/1025002}. As we encounter fairly extreme density and temperature contrasts in the evolution, we need to enforce an automated recovery procedure to ensure positivity throughout, and we do so by fixing the minimum pressure and density values equal to $10^{-14}$ and $10^{-12}$ code units respectively. We follow the current sheet evolution for up to 2550 s (42 minutes), and usually save data with 42.5 s cadence, which gives 60 simulation snapshots. We use periodic boundary conditions in the $x$, and open boundary conditions in the $y$ direction. The typical wall clock time of a single run is $\approx 20$ hours for parallel computation with 15 CPUs.   


\section{Results and Analysis}\label{results}

\subsection{Global evolution}\label{sub:glob_evol}

The spatial distribution of the absolute current density, $|J_z|$ is shown in Fig. \ref{fig:jzmap} for four different time stages, with plasma-$\beta=0.2$, and resistivity, $\eta=0.001$ (equivalent to $1.2 \times 10^{14}$ cm$^2$ s$^{-1}$ in physical unit). Fig. \ref{fig:jz_t00} represents the initial configuration of the current sheet, which is located around $y=0$ and extends all along the $x$ direction. Fig. \ref{fig:jz_t21}) and its animated view clearly shows that the current sheet narrows as a result of the thermodynamic evolution driven by radiative losses and quickly thereafter fragments, forming a pronounced chain of many small-scale plasmoids due to the combination of thermal and tearing instabilities. We calculate the force (per unit volume) along the $y$-direction, $\displaystyle{F_y = -\frac{\partial p }{\partial y}+B_x \bigg(\frac{\partial B_y}{\partial x}-\frac{\partial B_x}{\partial y}\bigg)}$. At $t=0$, the system maintains the equilibrium condition and hence $F_y=0$, but when the system evolves, the equilibrium is violated due the thermally influenced tearing mode instability, where we see $F_y$ has a dominating positive force (upward direction) for $y<0$, and negative (downward direction) dominating force for $y>0$ in the vicinity of the current sheet at $x=0$ between $y=\pm 0.75 \times 10^4$ km at $t=300$ and 429 s before the fragmentation stage of the current sheet (see Fig. \ref{fig:Fy}). Therefore, the current sheet is squeezed along the $y$-direction. As time progresses, the small islands coalesce and merge with each other at later times shown in Figs. \ref{fig:jz_t40} and \ref{fig:jz_t57}. The evolution of the plasma density, $\rho$, at the same stages is shown in Fig. \ref{fig:rhomap_wfl}. The initial configuration of the density distribution shown in Fig. \ref{fig:rhomap_t00} has an enhanced density region present in the vicinity of $y=0$ according to equation (\ref{eq:rho_in}). In the later stage, at $t=900$ s (Fig. \ref{fig:rhomap_t21}), the current sheet fragments into smaller plasma blobs, and these merge with each other at the later stages as shown in Fig. (\ref{fig:rhomap_t40}) and (\ref{fig:rhomap_t57}) respectively. The same instants but seen in the temperature evolution are shown in Fig. \ref{fig:Tmap}. Comparing Figs. \ref{fig:rhomap_t21}, \ref{fig:rhomap_t40}, \ref{fig:rhomap_t57} with the Figs. \ref{fig:Tmap_t21}, \ref{fig:Tmap_t40}, \ref{fig:Tmap_t57} shows that the temperature depletion regions are formed at the regions where the plasma materials are condensed. We calculate the radiative loss at $x=0$ between $y=\pm 1.25\times 10^4$ km, which covers the entire vertical domain of the current sheet. At $t=0$, the radiative loss is equal to the background heating, $H_{bgr}$, which is constant with time. However, when the system evolves due to the thermally influenced tearing mode, the radiative loss term dominates over the background heating within the selected region. This is shown in the Fig. \ref{fig:rc_ycut}, where the radiative loss dominates over $H_{bgr}$ at $t=429$ s, which leads to a temperature drop to $21000$ K within that region as shown in Fig. \ref{fig:T_ycut}. The spatial distribution of the density on a horizontal cut all along the current sheet ($y=0$) is shown in Fig. \ref{fig:rho_yo}, for two different evolution stages ($t=900$ and 2447 s). Due to the relative motions of the plasmoids along the $\pm x$ direction, they merge to form denser plasmoids, separated by density depletion regions, which is reflected in Fig. \ref{fig:rho_yo}. The temporal evolution of the instantaneous minimum and maximum density (\textit{left}) and temperature (\textit{right}) as computed over the entire domain are shown in Fig. \ref{fig:rho_T_evol}. These extrema are virtually always encountered along the central current sheet. It shows that the minimum density ($\rho_{min}$) falls gradually with the evolution, whereas the maximum density ($\rho_{max}$) increases with time, which means that the overall density contrast of the medium increases with time. It is estimated that the ratio of $\rho_{max}$ to $\rho_{min}$ changes by a factor of $\approx 10^1$ to $\approx 10^6$ between the initial and the final times. The equilibrium temperature of the initial setup is constant (0.5 MK) for the entire simulation domain, and the minimum temperature during the evolution is not allowed to drop below 1000 K, which is well above the minimal temperature of the exploited cooling curve. Hence, the maximum ($T_{max}$) to minimum ($T_{min}$) temperature ratio of the medium is unity at the initial stage, and rises up to the factor of $\approx 10^4$ at the final time. The local thermodynamic analysis for some typical, selected plasmoids are shown in Figs. \ref{fig:rho_zoom}, \ref{fig:temp_zoom} and \ref{fig:rc_zoom}. The zoomed version of a local plasmoid during the evolution stage for two different times, $t=900$ and $2447$ s are shown in Fig. \ref{fig:rho_zoom}(b) and \ref{fig:rho_zoom}(d) respectively, where the variation of the mass densities along the horizontal cuts are marked, and shown in Figs. \ref{fig:temp_zoom}(c) and  \ref{fig:temp_zoom}(f) respectively. The maximum density enhancement along the marked line for $t=900$ s is $\sim 10^{-13}$ g cm$^{-3}$ (Fig. \ref{fig:temp_zoom}(c)), whereas for $t=2447$ s, the value rises upto $\sim 10^{-12}$ g cm$^{-3}$ (Fig. \ref{fig:temp_zoom}(f)). This implies that the condensations formed by thermal instability within smaller-scale plasmoids get collected into larger, condensed regions, as the plasmoids merge by coalescence. Similarly, the temperature variation in the same zoomed region is shown in Fig. \ref{fig:temp_zoom}. It shows that the minimum temperature within this domain is $\sim 10^4$ K, and the temperature depletion regions correspond to the overdense regions of Figs. \ref{fig:rho_zoom}(a) and \ref{fig:rho_zoom}(c) respectively. The radiative loss for an optically thin medium, which depends on the local density and temperature ($\rho^2 \Lambda(T)$) is shown in Fig. \ref{fig:rc_zoom}. The variation of the radiative loss along the horizontal marked lines of Figs. \ref{fig:rc_zoom}(b) and \ref{fig:rc_zoom}(e) are shown in Figs. \ref{fig:rc_zoom}(c) and \ref{fig:rc_zoom}(f) respectively. It is to be noted from Figs. \ref{fig:temp_zoom}(b) and \ref{fig:rc_zoom}(b) (or from Figs. \ref{fig:temp_zoom}(e) and \ref{fig:rc_zoom}(e)) that the radiative loss of the temperature depleted region is more than its surrounding. This is because of the density enhancement of that region compared to its surroundings (see Fig. \ref{fig:rho_zoom}(b) or \ref{fig:rho_zoom}(d)), and the $\rho^2$ term dominates over the $\Lambda(T)$ resulting the increase of the radiative loss, $\rho^2 \Lambda(T)$. The condensations that are entrapped within the coalescing plasmoids thereby also show rapid variations of the radiative losses across their edges, much like the prominence corona transition region (PCTR), also at play for individual coronal rain blobs. Note that the precise variation of temperature, density and radiative losses is here not incorporating effects of anisotropic thermal conduction, and therefore we have extremely sharp transitions, as explained also in \citep{2021A&A...655A..36H}.

\subsection{Growth rate and scaling relation}\label{subsec:growth}
The velocity distribution of the plasma motion is shown in Fig. \ref{fig:vxmap}, where the magnitude is scaled with respect to the Alfv{\'e}n velocity $v_A$, which is measured based on the magnetic field strength, $B_0=2$ G, and mass density of the equilibrium current sheet, $\rho_c = 2.81 \times 10^{-15}$ g cm$^{-3}$. The Alfv{\'e}n time scale is measured by $t_A = \bar{L}/v_A$, where, $\bar{L}$ is the unit length of $10^9$ cm. It is evident from Figs. \ref{fig:vxmap_t21}, \ref{fig:vxmap_t40} and \ref{fig:vxmap_t57} that the velocity, $v_x$ stays localized in the vicinity of the current sheet ($y=0$). Note that in line with our initial single-island magnetic field perturbation, we see a pronounced rightward motion in the right half of the domain, and a leftward one at left. We later see typical Petschek-like signatures in the flow fields in between islands, e.g. especially about the middle $x=0$, with super-Alfv\'enic outflow speeds bounded by slow shocks. Fig. \ref{fig:vx_eta001} represents the evolutionary nature of a current sheet in an adiabatic and non-adiabatic conditions. It is evident from the figure that the instantaneous maximum velocity growth for the non-adiabatic case is more rapid than for the adiabatic conditions. The evolutionary behaviour of the current sheet configuration due to thermal and tearing instabilities is shown by the black curve in Fig. \ref{fig:vx_eta001} for plasma-$\beta=0.2$, and a given resistivity value, $\eta = 0.001$, while the evolution for different $\eta$ values are shown in Fig. \ref{fig:vx_alleta}. As a diagnostic measurement of the instability, we determine the evolution of the instantaneous maximum absolute velocity, $|v_x|_{max}$. From Fig. \ref{fig:vx_eta001}, it can be noticed that this evolution exhibits three distinct phases: (i) the early phase (between $t=0$ to 250 s), where the velocity growth occurs exponentially (linearly on the logarithmic-linear scale), which is called the linear growth regime, (ii) the next phase between $t=250$ to 665 s, where the growth rate is slower compared to the linear phase, which is called the Rutherford regime \citep{rutherford1903}, and (iii) the final phase, which we call the post-Rutherford regime, starts at $t=665$ s, where the instability suddenly develops in an explosive way, and finally saturates at a later time. To infer the evolution rates quantitatively for all the different phases, we calculate the growth rates by scaling it with respect to the Alfv{\'e}n time scale, $t_A$. We define the growth rate as, $\displaystyle{\gamma = {{\rm d}\big(\mathrm{ln}(|v_x|_{max})\big)}/{{\rm d}t}}$. To estimate the linear growth rate, $\gamma_{lin}$, we calculate the growth rate in the linear regime by taking the mean value of the slope, which gives, $\gamma_{lin}=3.76 \times 10^{-1} t_A^{-1}$. This value is larger by an order of magnitude as compared to the studies of the double current sheet problem \citep{1992PhFlB...4.3811O, 2011PhPl...18e2303Z, 2017PhPl...24h2116A, 2021PhPl...28h2903P}, where the radiative cooling effect (or other non-adiabatic effects, e.g. thermal conduction) is not incorporated. This implies that the larger linear growth rate can be ascribed to the non-adiabatic effects of the radiative cooling and background heating. This is also in agreement with our own study for a single current layer model reflected in Fig. \ref{fig:vx_eta001}, where the average growth rate for the adiabatic medium is smaller than the non-adiabatic case. Similarly, we estimate the average growth rates for the Rutherford regime ($\gamma_{Ruth}$) and the post-Rutherford regime ($\gamma_{PR}$) for different resistivity values within the range of $\eta= 0.0001$ to 0.005. The velocity evolution for some selected resistivity values are shown in Fig. \ref{fig:vx_alleta}. This shows that the explosive phase of the evolution starts at later times for higher resistivity values, and converges at the final stage. For a Sweet-Parker type current sheet (where the inverse aspect ratio of the current sheet follows the scaling relation, $l_s/L \sim S_L^{-1/2}$), the thickness of the current sheet increases with the resistivity \citep{2007PhPl...14j0703L}, which reduces the growth rate of the tearing mode when it is normalized with respect to the Alfv{\'e}n crossing time along the length of the current sheet (x-direction in our case). Hence, the explosive phase of the evolution in our simulation starts at later times for higher resistivity values. We have estimated the absolute current density, $|J_z|$ (normalized to unity) before the fragmentation stage of the current sheet ($t=214$ s) by taking a vertical cut along the $y$-direction at $x=0$ for two different resistivities, $\eta =$ 0.0001 and 0.001, to confirm that the thickness of the current sheet is increasing with resistivity (see Fig. \ref{fig:cs_width}). The resistivity dependence for the different evolution phases is shown in Fig. \ref{fig:gr_eta}. Fig. \ref{fig:ruth_gr} shows that $\gamma_{Ruth}$ follows a power-law dependence with the resistivity, $\gamma_{Ruth} \approx \eta^{-0.1}$ with a correlation coefficient (CC) of $-64.1$\%. The resistivity scaling relation for the post-Rutherford and the entire non-linear regimes are shown in Figs. \ref{fig:pruth_gr} and \ref{fig:avg_gr} respectively. We estimate the growth rate scaling relations for the post-Rutherford regime, $\gamma_{PR} \approx \eta^{0.03}$ (with CC = 59.9\%), and the entire non-linear regime, $\gamma_{avg} \approx \eta^{0.017}$ (with CC = 66.7\%). Previous studies by \cite{2011PhPl...18e2303Z, 2017PhPl...24h2116A, 2017PhPl...24c2115G} (and references therein) have reported the resistivity scaling relation of the non-linear growth rates for the DTM setup in the adiabatic environment, which have larger power-law indices compared to our estimation. Hence, our study infers that the resistivity dependence on the non-linear growth rates is weaker when the thermal instability reinforces the tearing mode.

\subsection{Plasmoid distribution and density filling factor}\label{subsec:plasmoids}

Due to the combined thermal and tearing instability, the current sheet becomes unstable and the magnetic islands are formed as shown in Figs. \ref{fig:jz_t40} and \ref{fig:jz_t57}. These islands show the coalescence tendency to merge with the neighbourhood companions to form larger plasmoids (see Figs. \ref{fig:rhomap_t40} and \ref{fig:rhomap_t57}). The mass density of the plasmoids is higher than in the local background medium. Since we wish to quantify some statistical properties on the evolving and coalescing plasmoids and their internal thermodynamics, we need a criterion to identify and count them. We define the plasmoids by a density threshold condition: if the density of a region is more than a density threshold, $\rho_{th}$, then we call it a plasmoid. Here, the threshold density, $\rho_{th}$ is defined as the 0.03\% of the peak density ($\rho_{max}$) for an instantaneous time. This means the $\rho_{th}$ varies with time, according to the temporal variation of $\rho_{max}$. This is equivalent to capturing the density regions up to $3\sigma$ level (99.97\%) of the peak density of a Gaussian distribution and separating out that density enhanced regions from their local background medium. We also calculate the fraction of volume occupied by these density enhanced structures (or plasmoids) with respect to the entire volume of the simulation domain (which is $256 \times 256$ Mm$^2$), which we call the density filling factor. The number distribution of the plasmoids and the density filling factor with time for different resistivities are shown in Fig. \ref{fig:plasma_ff_eta}. At the initial time ($t=0$), the ratio between $\rho_{min}$ to $\rho_{max}$ is 16.67\% (which is more than 0.03\%). Hence, the threshold density captures the entire volume of the simulation domain and therefore the plasmoid numbers obtained in this method remain unity unless the density contrast satisfies the density threshold condition as mentioned above (see Fig. \ref{fig:plasmanum_eta}). After this phase, when the current sheet fragments, the maximum number of small-scale plasmoids form, which is represented by the highest peak of the distribution curve of Fig. \ref{fig:plasmanum_eta}. We also notice from Fig. \ref{fig:plasmanum_eta} that the current sheet fragments into more plasmoids for lower resistivity values, and the fragmentation phase occurs at a later time for higher resistivity. We obtain an inverse relation of the maximum plasmoid numbers with resistivity that follows a scaling relation, $N \sim \eta^{0.223}$ (see Fig. \ref{fig:plasma_num_sl}). This is similar to the case, $N \sim \eta^{3/8}$ obtained by \cite{2007PhPl...14j0703L} for adiabatic medium, where it is reported that the current sheet thickness depends on the resistivity. Thus, our simulation results also imply that the current sheet thickness is a function of the resistivity similar to the Sweet-Parker type current sheet, which is also being reflected in Fig. \ref{fig:cs_width}. Similarly, the density filling factor for the plasmoids at the initial phase is unity as shown in Fig. \ref{fig:ff_eta}, and it decreases with time so the volume fraction occupied by the plasmoids diminishes at the later stage of the evolution. The distribution of the maximum plasmoid numbers with the Lundquist number, $S_L$ is shown in Fig. \ref{fig:plasma_num_sl}, which shows that the number scales with $S_L^{0.223}$ (with a correlation coefficient of the linear fit equal to 0.96). The stability analysis by \cite{2007PhPl...14j0703L} reports that this number distribution scales as $S_L^{0.375}$ in an adiabatic medium. This suggests that the dependence of the Lundquist number in the number distribution of secondary islands is weaker for a non-adiabatic medium. In the Sweet-Parker type current sheet, the inverse aspect ratio of the current sheet, $l_s/L$ follows the scaling as $l_s/L \sim S_L^{-\alpha}$, where $\alpha = 0.5$. In our study, for $l_s = 0.5, L=12.8$, and $S_L$ in the range between $4.67 \times 10^3$ to $2.34 \times 10^5$, we estimate the value of $\alpha$ that varies between 0.26 to 0.38, which are clearly lower than 1/2. This implies that the current sheet is thicker than the Sweet-Parker type within our explored domain of the Lundquist number. We also estimate the temporal variation for the distribution of plasmoid numbers and the density filling factor for different plasma-$\beta$ values that are shown in Figs. \ref{fig:plasmanum_beta} and \ref{fig:ff_beta} respectively. To vary the plasma-$\beta = 0.2, 0.1$ and 0.02, we tune the initial temperature, $T_i=0.5, 0.25$ and 0.05 MK respectively keeping the magnetic field strength, $B_0 = 2$ G as constant. From \ref{fig:plasmanum_beta} we see that the maximum fragmentation phase of the current sheet occurs at a later time for higher plasma-$\beta$ values, but there is no specific trend for the peak values of the plasmoids with plasma-$\beta$. We perform the same analysis by upgrading the maximum numerical resolution by $4096 \times 4096$ with $\eta=0.001$ and plasma-$\beta=0.2$, keeping all the other parameters constant (see Fig. \ref{fig:plasma_ff_reso}). The numerical resistivity of the medium decreases for higher resolution values, and hence the maximum plasmoid numbers in the fragmentation phase increase as shown in Fig. \ref{fig:plasmanum_reso}, though the results do not alter significantly for different resolutions. The result is also consistent with the study of the physical resistivity cases as shown in Fig. \ref{fig:plasmanum_eta}. We also notice that the variation of the density filling factor distribution with time is not significantly different for different plasma-$\beta$ (see Fig. \ref{fig:ff_beta}) and numerical resolutions (see Fig. \ref{fig:ff_reso}).

Finally, we estimate the kinetic energy density
\begin{align}\label{eq:ke}
    \mathrm{KE} = \frac{1}{V}\iint \frac{\rho v^2}{2}{\rm d}x {\rm d}y,
\end{align}
for each time step, where we integrate over of the full simulation domain, $V=l_x l_y$, and also quantify the maximum Ohmic heating, $H_{ohm} = \eta |J_z|_{max}^2$. The evolution of the kinetic energy density and the maximum Ohmic heating rate for different plasma-$\beta$ are shown in Fig. \ref{fig:ke} and \ref{fig:ohm_heat} respectively, which shows that the maximum energy dissipation per unit time by the Ohmic heating is approximately two orders of magnitude less than the kinetic energy. There is more rapid temporal variation in the later merging stages of the plasmoids as seen in the Ohmic heating extremal evolution, and only some modest dependency of the overall energetics on the plasma beta parameter.


\section{Discussion and summary}\label{discussion}

The instability problem addressed in this work can be related to the preflare current layer model. A cartoon geometry of the configuration for a current sheet associated with a preflare event is shown in Figure 5 of \cite{2021SoPh..296...74L}, where the theoretical linear stability analysis is carried out with the inclusion of viscosity, electrical and thermal conductivity, and radiative cooling. In the follow-up works by \cite{2021SoPh..296...93L, 2021SoPh..296..117L}, the effects of the guiding magnetic field and the oblique fragmentation of the current sheet are investigated in linear MHD with an analysis of the growth rate and spatial periodicity scales of the instability. Whereas in the present work, we extend the analysis for growth rate for both linear and non-linear domains with the incorporation of radiative losses, and constant background heating by a series of resistive nonlinear, high resolution MHD simulations. The instability that occurs in this work can be seen as a thermal instability enhanced pathway to rapid small-scale tearing behavior, where thereafter coalescing islands evolve and collect small condensations into larger entrapped cool plasma sites within magnetic islands. At the initial time ($t=0$), the thermal balance is maintained due to the equal and counteracting effects of the radiative loss and background heating. However, due to the magnetic field perturbation and the finite resistivity, some of the sections within the current sheet begin to increase the density and hence lose more heat due to radiation as compared to the constant background heat. Hence, the thermal imbalance occurs which leads to instability. As a result, the current sheet starts to disintegrate into the form of plasmoids and these move along the current sheet by merging with the neighboring plasmoids. The regions of main solar flare energy release occur from current sheet regions, which can lead to the outburst of a coronal mass ejection and modern observations may detect fine scale multi-thermal structure in the reconnecting sheet, by the brightening in the ultraviolet (UV) regime \citep{2016NatSR...624319J, 2021ApJ...920..102W}. As previously shown in Fig. \ref{fig:rhomap_wfl}, the secondary islands are present near the magnetic X-points in the form of plasmoids which appear during the explosive phase. The local Lundquist number defined by, $S_L=l_x v_A/\eta$ (where $l_x$ is the length of the current layer), in the series of our simulations is in the range of $4.67 \times 10^3$ to $2.34 \times 10^5$. We see that the formation of the plasmoids occurs over the entire explored Lundquist number range in our simulation. Note that this is extending the chaotic reconnection process to much lower Lundquist range than previously found from purely resistive (but otherwise adiabatic) evolutions. Indeed, in the previous studies by \cite{2007PhPl...14j0703L, 2009PhRvL.103j5004S, 2009PhPl...16k2102B, 2017PhPl...24h2116A}, it is reported that the plasmoid formation occurs only beyond a minimum threshold value of the Lundquist number, $S_L \sim 10^4$, where those studies were limited for adiabatic regimes. This threshold is also determined by the inhomogeneous inflow and outflow in the reconnecting sheet, which suppresses the growth of tearing in an adiabatic medium \citep{2018ApJ...859...83S}. This means that the formation of plasmoids may occur for lower Lundquist numbers ($\lesssim 10^4$) due to the non-adiabatic effects of thermal instability, and this lower Lundquist range is easily resolvable numerically. Still, extreme resolution is warranted because the fine-scale effects this time are due to the thermally unstable nature of the solar coronal plasma. We also investigate an experiment by switching off the physical resistivity (i.e. $\eta=0$) but keeping the adiabatic effects on for a sufficiently high numerical resolution of $2048 \times 2048$, so that this experiment is only evolving due to numerical (unresolved) resistivity. This experiment shows that even for a low (numerical) resistivity value, the formation of plasmoids occurs due to the fragmentation of the current layer, and we get explosive behavior. \\

We highlight the novel features and the key results of this work in the following.
\begin{enumerate}
    \item We set up a numerical experiment of a 2D current sheet model by incorporating non-adiabatic effects of radiative loss and constant background heat in a resistive MHD simulation using MPI-AMRVAC. Due to the magnetic field perturbations (Equations \ref{eq:delbx} and \ref{eq:delby}), the equilibrium of the system breaks down, and the instability kicks in, in the form of thermal and tearing modes. The current layer starts to disintegrate to form secondary islands, which move along the current sheet by merging with the neighboring plasmoids.
    
    \item The thermodynamical behavior of the current sheet region, and for a local plasmoid in particular is analyzed. We see the density enhancement due to the accumulation of the neighboring plasmoids, or density drop due to the disintegration of the current layer. Accordingly, the temperature of the medium drops for the density enhanced regions and rises up for the density depletion regions. We also estimate the global behavior of the energy loss in the medium due to the optically thin radiation. 
    
    \item We compute the growth rates for the linear and non-linear phases of the evolution and estimate the scaling relations with the resistivities for different non-linear phases. We find that the growth rate obtained from our model is faster by an order of magnitude, and weaker with resistivity in comparison with earlier works by \cite{1992PhFlB...4.3811O, 2011PhPl...18e2303Z, 2017PhPl...24h2116A, 2021PhPl...28h2903P}, where they have assumed the medium to be adiabatic. We notice the occurrence of the explosive nature of the evolution within the resistivity domain of $\eta = 5 \times 10^{-3}$ to $10^{-4}$ in our work. This is a regime where thermal instability enhanced fragmentation triggers small-scale tearing effects.
    
    \item The temporal variation of the generated plasmoid numbers and the associated density filling factors are estimated for different $\eta$ and plasma-$\beta$ that are relevant for the solar corona. We calculate the scaling relation of the maximum plasmoid numbers with the Lundquist number, $S_L$, and notice it to vary as $S_L^{0.223}$, which is smaller than the value estimated by \cite{2007PhPl...14j0703L} for an adiabatic medium. This implies a thicker tearing-unstable current sheet than the usual Sweet-Parker type, and indicates that the thermal instability facilitates the triggering of tearing modes. We also investigate the analysis for higher numerical resolution (4096 $\times$ 4096) and see that it does not alter the results significantly. 
    
    \item The time evolution of the kinetic energy density and Ohmic dissipation rate are calculated. The comparison between these two energies shows that the energy dissipation per unit time due to Ohmic heating is around two orders of magnitude less than the kinetic energy. The later stages show clear Petschek-like super-Alfv\'enic outflow regions in between the merged, larger islands.
\end{enumerate}

We performed a detailed magnetohydrodynamic study of a current sheet model liable to both thermal and tearing instabilities. We did not incorporate the effects due to thermal conduction, a guide magnetic field, or gravity in this model. A more realistic 3D model with the incorporation of these effects can be explored in near future. However, the current idealized study sheds new light on the formation mechanisms of plasmoids, and explosive reconnection behavior of a preflare current layer model, which is one of the important aspects of solar coronal heating. Our findings suggest that multi-thermal plasma aspects must be common in flaring regions, where both hot islands with their entrapped cooler condensations must show up cospatially in different wavebands.

\begin{acknowledgements}
Data visualization and analysis are performed using ParaView (\url{https://www.paraview.org}) and python (\url{https://www.python.org/}). SS and RK acknowledge support by the C1 project TRACESpace funded by KU Leuven. RK acknowledges the support by the European Research Council (ERC) under the European Unions Horizon 2020 research and innovation program (grant agreement No. 833251 PROMINENT ERC-ADG 2018) and a FWO project G0B4521N. The
computational resources and services used in this work were provided by the VSC (Flemish Supercomputer Center), funded by the Research Foundation Flanders (FWO) and the Flemish Government - department EWI. We thank N. Yadav, X. Li, and J. Hermans for useful discussions during the course of this work.  
\end{acknowledgements}

 \bibliographystyle{aa} 
 \bibliography{manuscript_final_twocol} 

\end{document}